\documentclass[fleqn,usenatbib]{mnras}

\usepackage{newtxtext,newtxmath}

\usepackage[T1]{fontenc}

\DeclareRobustCommand{\VAN}[3]{#2}
\let\VANthebibliography\thebibliography
\def\thebibliography{\DeclareRobustCommand{\VAN}[3]{##3}\VANthebibliography}

\usepackage{graphicx}	
\usepackage{amsmath}	
\usepackage{comment}
\usepackage{booktabs}  
\providecommand{\abs}[1]{\lvert#1\rvert}

\usepackage{caption}
\usepackage[flushleft]{threeparttable} 
\usepackage{xspace}

\newcommand{\msun}{M$_{\sun}$\xspace}
\newcommand{\rsun}{R$_{\sun}$}
\newcommand{\mearth}{M$_{\oplus}$}

\newcommand{\h}{^{\rm h}}
\newcommand{\m}{^{\rm m}}

\newcommand{\pmunit}{mas\,yr$^{-1}$}
\newcommand{\ergs}{\rm {erg} \, \rm{s}^{-1}}
\newcommand{\dev}{\mathrm{d}}
\newcommand{\Pobs}{$P_{\rm obs\ }$}

\newcommand{\Pdotobs}{$\dot{P}_{\rm obs}$}
\newcommand{\Pdotint}{$\dot{P}_{\rm int}$}
\newcommand{\PRESTO}{\texttt{PRESTO}}
\newcommand{\PULSARMINER}{\texttt{PULSAR\_MINER}}
\newcommand{\TEMPOTWO}{\texttt{TEMPO2}}
\newcommand{\DSPSR}{\texttt{DSPSR}}
\newcommand{\PDMP}{\texttt{PDMP}}

\title[Pulsars in NGC 6440]{Discoveries and Timing of Pulsars in NGC 6440}

\author[L. Vleeschower et al.]{\parbox{\textwidth}{
L.~Vleeschower,$^{1}$\thanks{E-mail: laila.vleeschowercalas@postgrad.manchester.ac.uk}
B.~W.~Stappers,$^{1}$
M.~Bailes,$^{2,3}$
E.~D.~Barr,$^{4}$
M.~Kramer,$^{4,1}$
S.~Ransom,$^{5}$
A.~Ridolfi,$^{4,6}$
V.~Venkatraman~Krishnan,$^{4}$
A.~Possenti,$^{6}$
M.~J.~Keith,$^{1}$
M.~Burgay,$^{6}$
P.~C.~C.~Freire,$^{4}$
R.~Spiewak,$^{1}$
D.~J.~Champion,$^{4}$
M.~C.~Bezuidenhout,$^{1}$
I.~C.~Ni\c{t}u,$^{1}$
W.~Chen,$^{4}$
A.~Parthasarathy,$^{4}$
M.~E.~DeCesar,${^7}$
S.~Buchner,${^8}$
I.~H.~Stairs,${^9}$
and J.~W.~T.~Hessels${^{10,11}}$
}
\\ \\ \\ \\
\parbox{\textwidth}{
$^{1}$Jodrell Bank Centre for Astrophysics, Department of Physics and Astronomy, The University of Manchester, Manchester M13 9PL, UK\\
$^{2}$  Centre for Astrophysics and Supercomputing, Swinburne University of Technology, P.O. Box 218, Hawthorn, VIC 3122, Australia\\
$^{3}$ ARC Centre of Excellence for Gravitational Wave Discovery (OzGrav)\\
$^{4}$Max-Planck-Institut f\"{u}r Radioastronomie, Auf dem H\"{u}gel 69, D-53121 Bonn, Germany\label{mpifr}\\
$^{5}$ National Radio Astronomy Observatory, 520 Edgemont Rd., Charlottesville, VA 22903, USA\\
$^{6}$INAF -- Osservatorio Astronomico di Cagliari, Via della Scienza 5, I-09047 Selargius (CA), Italy\\
$^{7}$ George Mason University, Fairfax, VA 22030, resident at the U.S. Naval Research Laboratory, Washington, D.C. 20375, USA \\
$^{8}$South African Radio Astronomy Observatory (SARAO), 2 Fir Street, Black River Park, Observatory, Cape Town, 7925, South Africa \\
$^{9}$Dept. of Physics and Astronomy, University of British Columbia, 6224 Agricultural Road, Vancouver, BC V6T 1Z1 Canada\\
$^{10}$ ASTRON, the Netherlands Institute for Radio Astronomy, Oude Hoogeveensedijk 4, 7991 PD Dwingeloo, The Netherlands \\
$^{11}$ Anton Pannekoek Institute for Astronomy, University of Amsterdam, Postbus 94249, 1090 GE Amsterdam, The Netherlands
}
}

\date{Accepted XXX. Received YYY; in original form ZZZ}

\pubyear{2022}

\begin{document}
\label{firstpage}
\pagerange{\pageref{firstpage}--\pageref{lastpage}}
\maketitle

\begin{abstract}
Using the MeerKAT radio telescope, a series of observations have been conducted to time the known pulsars and search for new pulsars in the globular cluster NGC 6440. As a result, two pulsars have been discovered, NGC 6440G and NGC 6440H, one of which is isolated and the other a non-eclipsing (at frequencies above 962 MHz) ``Black Widow'', with a very low mass companion (M$_{\rm{c}}$ > 0.006 \msun). It joins the other binary pulsars discovered so far in this cluster which all have low companion masses (M$_{\rm{c}}$ < 0.30 \msun). We present the results of long-term timing solutions obtained using data from both Green Bank and MeerKAT telescopes for these two new pulsars and an analysis of the pulsars NGC 6440C and NGC 6440D. For the isolated pulsar NGC 6440C, we searched for planets using a Markov Chain Monte Carlo technique. We find evidence for significant unmodelled variations but they cannot be well modelled as planets nor as part of a power-law red-noise process. Studies of the eclipses of the ``Redback'' pulsar NGC 6440D at two different frequency bands reveal a frequency dependence with longer and asymmetric eclipses at lower frequencies (962-1283 MHz).

\end{abstract}





\section{Introduction}

Globular clusters (GCs) are known to be ideal places for the production of exotic compact objects and binaries due to the high stellar densities in their cores ($\sim$ 10$^{5-6}$\,\msun pc$^{-3}$, \citealp[e.g.][]{Baumgardt+Hilker2018}), which result in dynamical interactions between the stars. The exchange interactions and collisions which result in the creation or disruption of binary systems, enable the formation of, for example, cataclysmic variables or low-mass X-ray binaries (LMXBs). In fact, early observations showed that the population of LMXBs per unit mass in GCs is orders of magnitude higher than in the Galactic disk \citep{Katz1975,Clark1975}. LMXBs are the progenitors of millisecond pulsars (MSPs) through the recycling model \citep{Alpar+1982, Radhakrishnan_Srinivasan1982}, in which accretion on to the neutron star (NS) spins it up to millisecond periods.

The large number of LMXBs in GCs, and the detection of radio sources in the  imaging of the latter, motivated surveys for radio millisecond pulsars. After the first discovery in 1987, M28A (PSR B1821--24A) by \citet{Lyne+1987}, 34 pulsars were soon discovered in 13 GCs which was followed by a second burst of discoveries in the 2000's thanks primarily to the Arecibo, Green Bank and Parkes telescopes (see Figure 1 from \citealt{Ransom2008}) and 150 pulsars were known in 28 clusters by 2018.
We have now entered a new phase. After the first searches in 2019 by the Five-hundred-meter Aperture Spherical Telescope \citep[FAST, e.g.][]{Pan+2021}
and the high gain/low system temperature MeerKAT telescope \citep{Ridolfi+2021}, the number of GC pulsars has increased to 236 in at least 36 different GCs\footnote{See \url{https://www3.mpifr-bonn.mpg.de/staff/pfreire/GCpsr.html} for the most up-to-date GC pulsar catalogue.} up to the end of February 2022. Thirty-two of those new pulsars have been discovered with FAST\footnote{Visit the FAST GC survey https://fast.bao.ac.cn/cms/article/65/} \citep[e.g.][]{Pan+2020, Wang+2020, Pan+2021}. 
Both FAST and MeerKAT, one in the northern hemisphere and the other in the southern hemisphere, respectively, have higher sensitivity than other telescopes that have recently been used for pulsars searches in GCs. FAST, in the $-15^\circ$ to $+65^\circ$ declination range of the sky, has been providing a factor of 2 to 3 times better raw sensitivity than the 305m Arecibo radio telescope \citep{Wang+2020}, while MeerKAT, the precursor of the Square Kilometre Array - SKA1-mid \citep{Dewdney+2009}, with a declination limit of $+44\degr$ \citep{Camilo+2018} is at least 3 times more sensitive than the systems used by the Parkes 64\,m telescope \citep{Stappers+Kramer2016,Bailes+2020}. 

A number of GCs are being searched using MeerKAT under the two Large Survey Projects: MeerTime\footnote{\url{http://www.meertime.org}} \citep{Bailes+2020} and TRAPUM \footnote{\url{http://www.trapum.org}} (TRAnsients and PUlsars with MeerKAT, \citealt{Stappers+Kramer2016}), both projects collaborate together to time and search for pulsars in GCs. These searches have proven to be fruitful with the discovery of 38 new pulsars in 11 different GCs so far\footnote{See \url{http://www.trapum.org/discoveries.html}} \citep[e.g.][]{Ridolfi+2021,Douglas+2022,Ridolfi+2022}.

NGC 6440 is a GC located at $\alpha = 17\h48\m52\fs84$, $\delta = -20\degr21\arcmin37\farcs5$ \citep{Pallanca+2021}, in the direction towards the bulge of the Galaxy at $l = 7.729\degr$ and $b = 3.800\degr$ in the constellation of Sagittarius. It is situated at a distance of 8.3 $\pm$ 0.4\,kpc from the Sun \citep{Pallanca+2021} and is moderately concentrated (with a central concentration $c = 1.86$, where $c = \log(r_{\rm t}/r_{\rm c})$, $r_{\rm t}$ and $r_{\rm c}$ are the tidal and core radii of the cluster, respectively) but apparently has no post-collapse core \citep{Trager+1993}. Integrated photometry suggests that this GC was among the most metal-rich GCs in the Galaxy \citep{Ortolani+1994}, with its metallicity comparable to that of the Sun ([Fe/H] $\sim$ -- 0.56, \citealt{Origlia+2008}), and  it has a reddening of $E(B-V)=1.15$ \citep{Valenti+2007}. NGC 6440 is particularly massive ($M > 4.42$ $\times$ 10$^5$\,\msun, \citealt{Baumgardt+Hilker2018}) and dense with a core and half-mass radius of $r_{\rm c}=0.26$\,pc and $r_{\rm h}=2.02$\,pc, respectively \citep{Pallanca+2021}, which corresponds to $r_{\rm c}=0.11$\,arcmin and $r_{\rm h}=0.84$\,arcmin. 

NGC 6440 is one of the richest X-ray clusters studied so far, only surpassed by Terzan 5 \citep{Bogdanov+2021}, with 25 X-ray sources within two core radii from its centre, identified by \citet{Pooley+2002, Heinke+2010} using Chandra images. One of them (CX1) was later identified as a transient luminous LMXB (SAX J1748.9$-$2021) with a rotation frequency of 442 Hz \citep{Gavriil+2006,Gavriil+2007}. Three years later, a second transient LMXB (NGC 6440 X$-$2) was discovered in the cluster with coherent 206 Hz pulsations \citep{Heinke+2010}, identified afterwards as an ultracompact accreting millisecond X-ray pulsar \citep[AMXP,][]{Altamirano+2010}. Furthermore, the cluster is known to host six pulsars (three isolated and three binaries), with dispersion measures (DMs) between 219.4 and 227.0\,pc\,cm$^{-3}$ \citep{Freire+2008}. Because of these characteristics, this cluster was selected as one of the high-priority targets in the Large Survey Proposals of MeerTime and TRAPUM. 

The six previously known pulsars are NGC 6440A \citep{Lyne+1996} and NGC 6440B-F \citep{Freire+2008}. NGC 6440A was found to be an isolated pulsar with a period of 289 ms, which is unusually long for a pulsar in a GC \citep{Lyne+1996} as only $\sim 4$\% of the pulsars in GCs have periods $P>100$\,ms. Almost 20 years later, 5 more pulsars were discovered using the S-band receiver (1650--2250 MHz) of the Green Bank Telescope (GBT). The other two isolated pulsars NGC 6440C and NGC 6440E have periods of 6.22 ms and 16.26 ms, respectively. 
NGC 6440B is a 16.7 ms pulsar in an eccentric binary system ($e = 0.57$) with an orbital period of $P_{\rm b} = 20.5$\,days.
NGC 6440D is a Redback pulsar, i.e. an eclipsing low-mass binary, with a spin period $P =$ 13.49\,ms and an orbital period $P_{\rm b}=0.28$\,days. It is the farthest from the GC centre as projected on the plane of the sky probably due to the formation in an exchange encounter \citep[see][]{Freire+2008}. Finally, NGC 6440F with a spin period of 3.79\,ms, is at the lower end of the spin period distribution of the GC. It has an orbital period  $P_{\rm b}=9.8$\,days and has a white dwarf companion. 

One of the most important results from \citet{Freire+2008} was the measurement of the rate of advance of periastron for NGC 6440B: $\dot{\omega} = 0.00391(18)\degr$\,yr$^{-1}$. Assuming that the latter is fully relativistic, this implies a total mass of $2.92 \pm 0.20$\,\msun \citep{Freire+2008}. For an edge-on inclination ($90\degr$) of the orbit this gives a mass of the pulsar of $M_{\rm p} = 2.3$\,\msun. If the high inclination were confirmed, it would make it the most massive NS known so far. 

In this paper we present the discovery of two more pulsars in this cluster and a study of pulsars NGC 6440C and NGC 6440D. It is structured as follows: the observations and data reduction are described in Section \ref{s:Obs and Data}. We report the discoveries in Section \ref{s:Discoveries}. The timing solutions of the pulsars discussed in this work are presented in Section \ref{s:Timing}. In Section \ref{s:Discussion} the results and their implications are summarised. Finally, the conclusions are presented in Section \ref{s:Conclusions}.



\section{Observations and Data Reduction}
\label{s:Obs and Data}

The possibility of a very massive NS for NGC 6440B was one of the motivations for its subsequent timing \citep[see][]{Kramer+2021}.
Previous observations of this pulsar made with the GBT greatly improved the precision of the measurement of $\dot{\omega}$. However, it has not been possible to obtain a significant detection of the Shapiro delay in the binary system, which suggests either a low orbital inclination or that the timing precision obtained with the GBT data was not high enough for a detection. This means that, using the GBT data alone, it was not possible to determine the individual masses of the components of the system.

Making use of the superior sensitivity of MeerKAT, a dense orbital campaign was mounted for this system with the main aim of detecting the Shapiro delay of NGC 6440B. This campaign consisted of 33 observations using the L-band (856-1712\,MHz) receivers of the MeerKAT telescope (with a central frequency $f_{\rm c} = 1284$\,MHz) under the MeerTime project pointing towards NGC 6440B to densely sample the $\sim20.5$-day orbit for two full revolutions. The number of antennas used for the observations depended on the orbital phase of the pulsar, with observations close to superior conjunction and periastron performed with the full array while other observations used just the central core of the array (see Table \ref{t:obs_summary}). This was to make sure that a number of observations had wide enough field-of-view (FOV) to cover a significant fraction of the cluster for commensal science, given the massive time investment. Each semi-axis of the single tied-array beam (using the full array of the telescope) has a minimum size of $\sim 0.1$\,arcmin at the central frequency $f_{\rm c} = 1284$\,MHz, covering the core radius of the GC.

We used the Pulsar Timing User Supplied Equipment \citep[PTUSE,][]{Bailes+2020} for data acquisition. Most observations had simultaneous data acquisition on two machines, one folding NGC 6440B at the topocentric period of the pulsar and recording pulsar timing archives, while the other recording full Stokes, PSRFITS format \citep{Hotan+2004} search mode filterbank data, coherently dedispersed at a DM of 220.922\,pc\,cm$^{-3}$, with $768 \times 0.42$\,MHz-wide frequency channels, and a time resolution of $\sim 9.5\,\mu$s. The fold mode data were used for the timing campaign whose results will be reported elsewhere (Venkatraman Krishnan et al. in prep.). The results presented in this paper used the search mode data to investigate two of the known pulsars in the GC and also to search for new pulsars. Apart from the observations performed for the timing campaign, we also analysed two other 30-minute test observations taken in April 2019 with a similar set-up as above. Table \ref{t:obs_summary} shows the list of observations employed in the present work and includes the observing dates, the central observing frequencies $f_{\rm c}$, the bandwidths (BW), the observation lengths $T_{\rm obs}$, the known pulsars found in the search, and the number of antennas used for each of the observations. It also includes information about the detections of the two new pulsars discovered as a result of this work. 

Around 75\% of the PTUSE data were initially analysed using the full frequency resolution and a time resolution of 76\,$\mu$s (HFR). Part way through the analysis it was realised that the data volume and processing could be sped up by reducing the number of channels to 384 (LFR). This resulted in no increase in dispersion smearing because of the initial coherent dedispersion of the data described above. The analysis was completed with the LFR data set.

All the HFR data were analysed following a typical acceleration search method. We first used the \texttt{rfifind} routine from \PRESTO~\citep{Ransom+2002} software suite to generate time-frequency masks and remove strong narrow-band and transient RFI from the data before searching. Around $\sim 15$\% of the data were masked for most of the observations. We then used \texttt{prepsubband} to generate 26 de-dispersed time series from a DM of 217.00\,pc\,cm$^{-3}$ in steps of 0.5\,pc\,cm$^{-3}$. The DM range was determined by considering $\pm 2.5$\,pc\,cm$^{-3}$ beyond the minimum and maximum DMs of the known pulsars in the cluster, while the DM step size was determined in order to have small DM smearing caused by an incorrect DM and sensible requirements for the processing.

A Fast Fourier Transform (FFT) was applied to each of the de-dispersed time series to obtain the power spectra; the red noise was removed from each of them using \texttt{rednoise}. Finally, the searches were carried out using \texttt{accelsearch} on the spectra, first without acceleration search ($z_{\rm max}=0$) and then with a $z_{\rm max}$ value of 200 (up to 8 harmonics), where $z=T_{\rm obs}^2a_l/(cP)$ is the number of Fourier bins drifted in the power spectrum (i.e. due to orbital motion) over the course of the observation, $a_l$ is the line-of-sight acceleration due to an orbital motion, and $c$ is the speed of light. The initial candidates were sifted using the code \texttt{accel\_sift.py}\footnote{\url{https://github.com/scottransom/presto/blob/master/examplescripts/ACCEL_sift.py}}. The initial parameters resulting from the search for each of the candidates were used to re-fold the raw data using \texttt{prepfold}. The results were then visually inspected. We will refer to this as the \PRESTO~search. 

The \PULSARMINER\ pipeline\footnote{\url{https://github.com/alex88ridolfi/PULSAR_MINER}} (v.1.5, see further details in \citealt{Ridolfi+2021}) was later implemented for the searches of all the LFR observations. \PULSARMINER\ automates and wraps all the processes from \PRESTO's accelsearch together to facilitate the tracking of all the processes and results of the system processing, with the help of GPU acceleration (\texttt{PRESTO\_ON\_GPU})\footnote{\url{https://github.com/jintaoluo/presto_on_gpu}}. This allowed us to include a more refined search, using a DM step size of 0.05\,pc\,cm$^{-3}$. We will refer to this as the \PULSARMINER~search. 

The acceleration search performed by \PRESTO~assumes that during an observation the pulsar has a constant acceleration along the line of sight. For this reason, the algorithm might fail to find pulsars with binary orbital periods ($P_{\rm b}$) shorter than $\sim 10$ times the duration of the observation, assuming a circular orbit \citep{Ransom+2003}. To be sensitive to orbital periods as short as $\sim 2.5$\,hrs, we performed a ``segmented search'', splitting all the observations (where possible) into sections of 60, 30 and 15 minutes, each of them searched individually. 

The discoveries presented below motivated us to observe the cluster as 
part of more extensive TRAPUM searches, as the latter is capable of generating more coherent beams and so can cover more of the cluster with the full MeerKAT sensitivity. These data enabled us to better localise the two newly discovered pulsars using a close tiling of the TRAPUM beams. Covering an area of $\sim 2$\,arcmin in radius, 288 coherent beams were synthesised, centred on the nominal cluster centre and using 60 antennas with a $> 70$\% overlap of the synthesised beams at 1284\,MHz \citep{Chen+2021}. The observations had an integration time of 4 hrs and the data were recorded as filterbanks, with a central frequency of 1284\,MHz and total intensity formed from the two orthogonal polarisations, 856\,MHz of bandwidth divided into 4096 frequency channels, with a time resolution of 76\,$\mu$s. The search of these data for new pulsars is still ongoing and will be presented in a future publication. Additionally, GBT data from both the Spigot \citep{Kaplan+2005} and GUPPI \citep{2007SPIE_DuPlain+} backends with the L-band and S-band receivers, respectively, were used to detect the newly discovered pulsars and, hence, to obtain a long-term timing solution (see Section \ref{s:Timing}). Full details of this GBT data set can be found in Ransom et al. (in prep).

\begin{table*}
\caption{Summary of the MeerKAT observations used in this analysis.}
\label{t:obs_summary}
\centering
\resizebox{\textwidth}{!}{
\begin{tabular}{lcccccccc}
\hline \hline 
\multicolumn{1}{l}{Date}   & \multicolumn{1}{c}{Start MJD} & \multicolumn{1}{c}{f$_{\rm c}$}  & \multicolumn{1}{c}{BW} & \multicolumn{1}{c}{T$_{\rm obs}$}  & \multicolumn{1}{c}{Known PSR}  & \multicolumn{1}{c}{G Detected}  & \multicolumn{1}{c}{H Detected} & \multicolumn{1}{c}{Number of}     \\ 
(yy-MM-dd-HH:mm) & & (MHz) & (MHz) & (min) & & Y/N & Y/N & Antennas\\
\hline 
19-03-04-03:10$^*$ & 58576.13  & 1283 & 321 & 30 & B & N & N & \\
19-03-04-03:10$^*$ & 58576.13  & 1444 & 321 & 30 & AB & N & N \\
19-07-18-18:32 & 58682.77  & 1284 & 642 & 120 & ABCEF & Y & Y & 57 \\
19-07-18-20:47 & 58682.86  & 1284 & 642 & 25 & ABCDE & N & Y & 57 \\
19-07-26-17:38 & 58690.73  & 1284 & 642 & 112 & ABDEF & Y & Y & 41 \\
19-07-28-17:02 & 58692.70  & 1284 & 642 & 120 & ABCDEF & Y & Y & 59 \\
19-07-28-19:17 & 58692.80  & 1284 & 642 & 120 & ABCDEF & Y & Y & 59 \\
19-07-30-16:42 & 58694.69  & 1284 & 642 & 164 & ABCDEF & Y & Y & 59 \\
19-07-30-20:30 & 58694.85  & 1284 & 642 & 163 & ABCDEF & Y & Y & 59 \\
19-08-01-15:46 & 58696.65  & 1284 & 642 & 120 & ABCDEF & Y & Y & 58 \\
19-08-01-18:14 & 58696.75  & 1284 & 642 & 120 & ABCDEF & Y & Y & 58 \\
19-08-03-16:53 & 58698.70  & 1284 & 642 & 120 & ABCDEF & Y & Y & 58 \\
19-08-03-19:12 & 58698.80  & 1284 & 642 & 120 & ABCDEF & \,\,\,Y$^s$ & Y & 58 \\
19-08-05-20:46 & 58700.86  & 1284 & 642 & 120 & ABCDEF & Y & Y & 58 \\
19-08-05-22:59 & 58700.95  & 1284 & 642 & 120 & ABCDEF & Y & \,\,\,Y$^s$ & 58 \\
19-08-07-18:26 & 58702.76  & 1284 & 642 & 120 & ABCDEF & Y & \,\,\,Y$^s$ & 42 \\
19-08-09-17:18 & 58704.72  & 1284 & 642 & 94 & ABCDEF & Y & Y & 41 \\
19-08-09-19:47 & 58704.82  & 1284 & 642 & 19 & ABCDEF & N & Y & 41 \\
19-08-11-16:48 & 58706.70  & 1284 & 642 & 120 & ABCDEF & Y & Y & 41 \\
19-08-11-19:22 & 58706.80  & 1284 & 642 & 120 & ABCDEF & \,\,\,Y$^s$ & Y & 41 \\
19-08-15-14:20 & 58710.59  & 1284 & 642 & 120 & ABCDEF & Y & Y & 38 \\
19-08-15-16:41 & 58710.69  & 1284 & 642 & 120 & ABCDEF & Y & \,\,\,Y$^s$ & 38 \\
19-08-17-16:39 & 58712.69  & 1284 & 642 & 120 & ABCDEF & Y & Y & 60 \\
19-08-17-19:11 & 58712.79  & 1284 & 642 & 120 & ABCDEF & Y & Y & 59 \\
19-08-19-15:52 & 58714.66  & 1284 & 642 & 120 & ABCDEF & Y & \,\,\,Y$^s$ & 63 \\
19-08-19-18:11 & 58714.66  & 1284 & 642 & 120 & ABCDEF & Y & Y & 63 \\
19-08-20-13:13 & 58715.55  & 1284 & 642 & 120 & ABCDEF & \,\,\,Y$^s$ & \,\,\,Y$^s$ & 63 \\
19-08-20-15:13 & 58715.63  & 1284 & 642 & 120 & ABCDEF & Y & \,\,\,Y$^s$ & 60 \\
19-08-20-17:37 & 58715.73  & 1284 & 642 & 120 & ABCDEF & \,\,\,Y$^s$ & Y & 60 \\
19-08-20-19:37 & 58715.81  & 1284 & 642 & 120 & ABCDEF & \,\,\,Y$^s$ & Y & 60 \\
19-08-21-16:52 & 58716.70  & 1284 & 642 & 120 & ABCDEF & Y & Y & 59 \\
19-08-25-16:43 & 58720.69  & 1284 & 642 & 180 & ABCDEF & \,\,\,Y$^s$ & Y & 61 \\
19-09-06-12:14 & 58732.50  & 1284 & 642 & 120 & ABCDEF & Y & \,\,\,Y$^s$ & 60 \\
19-09-06-14:38 & 58732.61  & 1284 & 642 & 120 & ABCDEF & Y & Y & 59 \\
20-11-14-14:05$^t$ & 59167.58 & 1284 & 856 & 90 & ABCDEF & Y & Y & 60 \\
20-11-14-15:42$^t$ & 59167.65 & 1284 & 856 & 143 & ABCDEF & Y & Y & 60 \\
20-12-01-13:06$^t$ & 59184.54 & 1284 & 856 & 240 & ABCDEF & Y & Y & 56 \\
\hline \hline 
\multicolumn{8}{l}{$^*$ test observation.}\\ 
\multicolumn{8}{l}{$^s$ pulsar found in the search.}\\ 
\multicolumn{8}{l}{$^t$ TRAPUM observations that were used for the timing analysis of both discovered pulsars.}
\end{tabular}
}
\end{table*}



\section{Discoveries}
\label{s:Discoveries}

The two new MSPs discovered in NGC 6440 are described in the remainder of this section.

\subsection{NGC 6440G}

NGC 6440G (PSR J1748--2021G) is an isolated pulsar, with a spin period of 5.22\,ms that was first found in the HFR data from UTC 2019-08-25-16:43 in the \PRESTO~search at a DM of 219.719\,pc\,cm$^{-3}$ and with no acceleration ($z =0$). Making use of the \PULSARMINER~search, the pulsar was subsequently found in five more observations (see Table \ref{t:obs_summary}). We used the best estimated period from the discoveries to fold all 33 MeerTime observations using \DSPSR\footnote{\url{http://dspsr.sourceforge.net}} and then optimised the period and DM using \PDMP\ from the \texttt{PSRCHIVE}\footnote{\url{http://psrchive.sourceforge.net/index.shtml}} package \citep{Hotan+2004,vanStraten+2012}. This resulted in the detection of the pulsar in 30 out of the 33 observations 
that were used for the searching in this work. The nondetections are due to the short duration of the remaining observations (see Table \ref{t:obs_summary}).

\subsection{NGC 6440H}

NGC 6440H (PSR J1748--2021H) was identified as a 2.85\,ms candidate in the HFR data set in the observation on 2019-08-05-22:59, but was not confirmed until we made detections in the LFR data. The non-zero acceleration and some pulse smearing observed in the diagnostic plots indicated the possibility of binary motion. We used the Period-Acceleration Diagram method (see \citealp{2001MNRAS_F}) to obtain starting estimates of the orbital parameters. First, we measured the barycentric observed spin period \Pobs and observed spin period derivative $\dot{P}_{\rm obs}$ from the initial search detections. The observations with high-signal-to-noise (S/N) detections were split into two, giving us a total of six different measurements of \Pobs and $\dot{P}_{\rm obs}$. The $\dot{P}_{\rm obs}$ was then converted into the line-of-sight acceleration $a_l = c(\dot{P}/P)_{\rm obs}$. The ($P_{\rm obs}$, $a_l$) points followed an elliptical curve indicating a circular orbit with orbital period of $P_{\rm b} \simeq 8.64$\,hours and a projected semi-major axis of only $a_{\rm p} \simeq 0.025$\,lt-s\footnote{Values obtained using the code \url{https://github.com/lailavc/circorbit}}.

Since we have closely-spaced detections, we used \texttt{fit\_circular\_orbit.py}\footnote{\url{https://github.com/kernsuite-debian/presto/blob/master/bin/fit_circular_orbit.py}} from \PRESTO~to improve the first guess values from the Period-Acceleration method. This code fits the observed spin period as a function of time, $P_{\rm obs}(t)$, returning improved orbital parameters. The results from this fit were an orbital period of $P_{\rm b} \simeq 8.66$\,hours and a projected semi-major axis of $a_{\rm p} \simeq 0.02541$\,lt-s. Those values were then further refined by phase-connecting the pulse Times of Arrival (ToA). The results of the timing are presented in the next section.



\section{Timing}
\label{s:Timing}

In this section we report the results of the timing analysis of the newly discovered pulsars as well as of NGC 6440C and NGC 6440D using data from both MeerKAT and GBT. For the case of the previously known pulsars, the MeerKAT search mode data were folded with \texttt{DSPSR}\ \citep{vanStraten+2011} using the then best ephemeris obtained from the analysis of several years of data from the GBT. For the new discoveries we initially used an ephemeris that included the barycentric spin period and DM of the best detection obtained with \texttt{prepfold}, and in the case of NGC 6440H, the orbital parameters from \texttt{fit\_circular\_orbit.py}. We then used \texttt{pat} from the \texttt{PSRCHIVE} package to extract the topocentric ToAs (from every archive where we detected the pulsar) by cross-correlating the pulse profiles against a noise-less template, built by fitting von Mises functions (using \texttt{paas} from \texttt{PSRCHIVE}) to the best-detection profile or one formed by adding the observations with detections, to form a mean pulse profile with higher S/N ratio. The ToAs were then referred to the Solar System Barycentre and fitted for different timing model parameters (e.g. celestial coordinates, spin parameters and the orbital parameters in the case of the binary systems) using \texttt{TEMPO2}\footnote{\url{https://bitbucket.org/psrsoft/tempo2/src/master/}} \citep{Hobbs+2006}. The initial timing solutions for the two newly discovered pulsars were then used to fold data from the GBT to potentially extend the timing baseline to more than 14 years. 

Before 2009, the GBT data included timing observations taken using the Spigot backend \citep{Kaplan+2005} using the S-band receiver (with usable band $\sim1.6$-$2.2$\,GHz). More details about these early observations can be found in \citet{Freire+2008}. NGC 6440 was later observed using the Green Bank Ultimate Pulsar Processing Instrument \citep[GUPPI;][]{DuPlain+2008}. These observations were made with both the L-band (i.e. 1.1-1.9\,GHz) and S-band (i.e. 1.6-2.4 GHz, with approximately the top $\sim 0.7$\,GHz usable) receivers. 

\begin{figure}
\centering
	\includegraphics[width=\columnwidth]{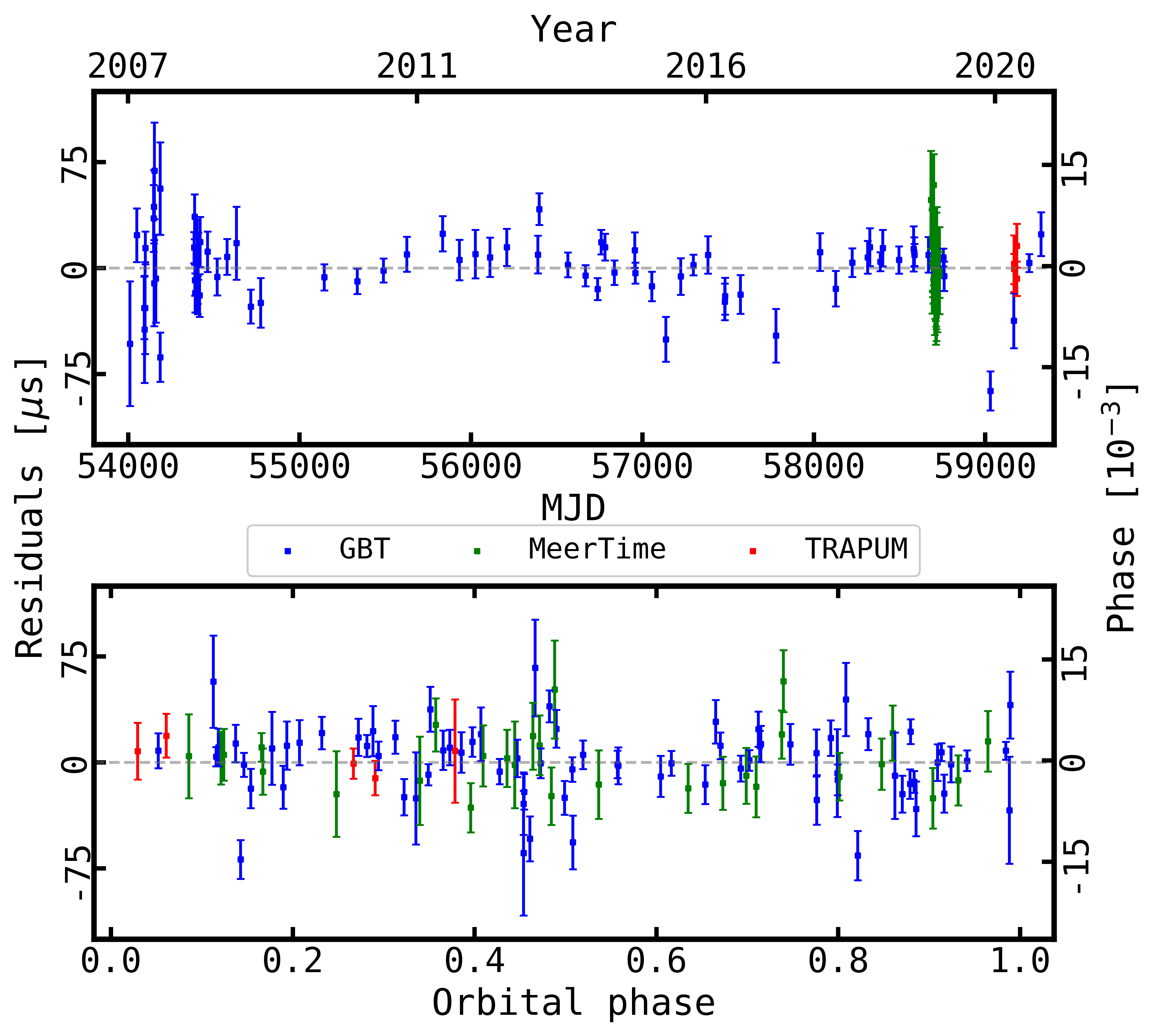}
    \caption{Timing residuals of NGC 6440H as a function of time (top) and orbital phase (bottom) obtained using GBT and MeerKAT data. The blue points indicate times of arrival from the GBT, while the green and red points show the times of arrival from MeerTime and TRAPUM, respectively.}
        \label{fig:NGC6440H_timingsol}
\end{figure}

\begin{figure}
\centering
	\includegraphics[width=\columnwidth]{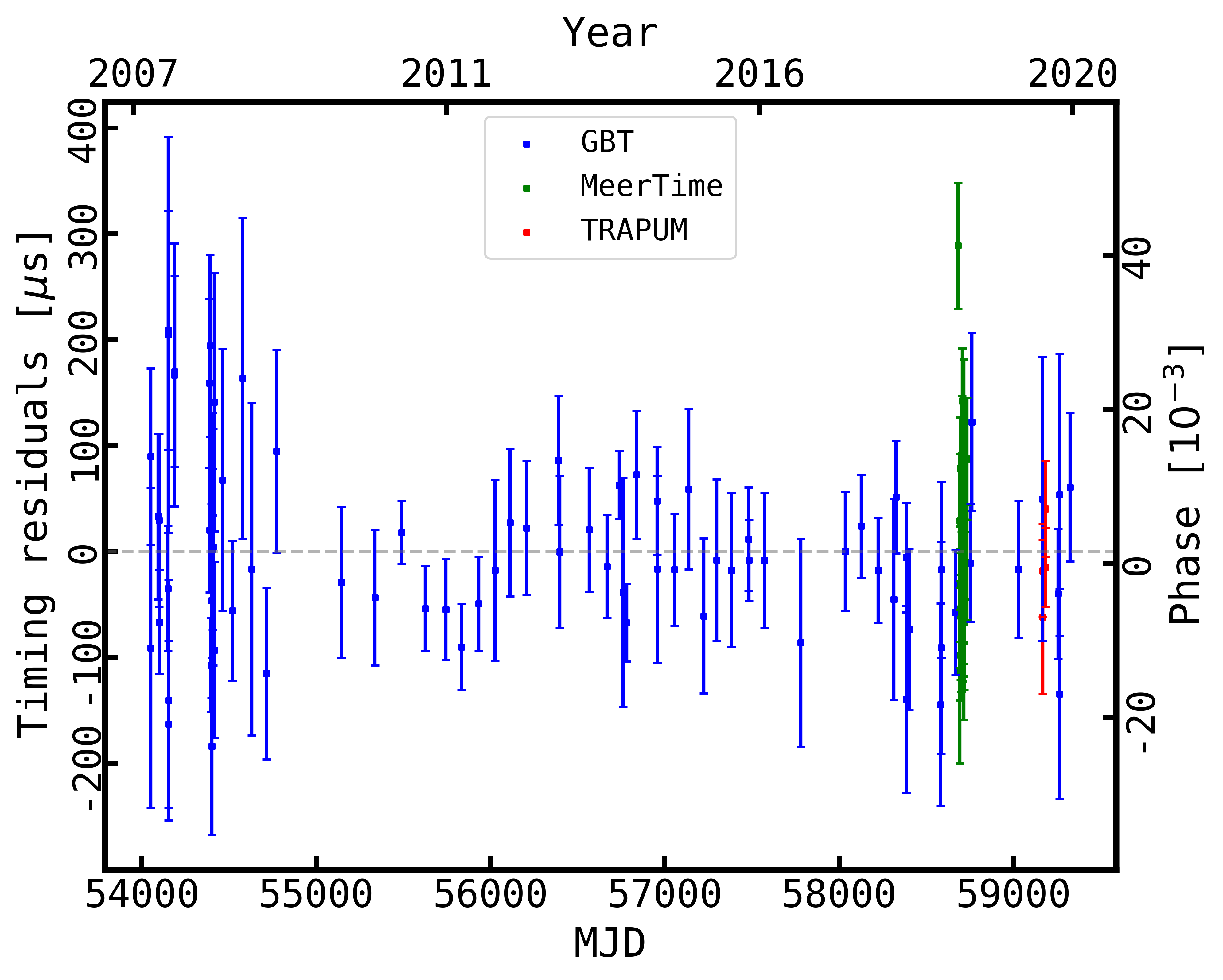}
    \caption{Timing residuals of NGC 6440G as a function of time (MJD) for the GBT and MeerKAT data.}
        \label{fig:NGC6440G_timingsol_MKT}
\end{figure}

Folding the GBT data with the best ephemeris obtained from the MeerKAT data resulted in 79 detections of NGC 6440G and 77 detections of NGC 6440H. As a result, we could obtain a phase-connected timing solution extending over more than 14 years (see Figures~\ref{fig:NGC6440H_timingsol} and \ref{fig:NGC6440G_timingsol_MKT}). For the case of NGC 6440G, the long-term timing solution benefitted from the localisation obtained using SeeKAT (see Section ~\ref{ss:Localisation}). All the timing properties (position, proper motions, binary parameters where applicable, etc.) for the two new pulsars are presented in Table~\ref{tab:timing_fitresults}. 

Furthermore, the GBT ToAs for NGC 6440C were combined with those from MeerKAT for further analysis and we used only MeerKAT TOAs for NGC 6440D. More details for each of the pulsars is provided in the next subsections.

\begin{table*}
\setlength\tabcolsep{10pt}
  \begin{threeparttable}
    \caption{Timing parameters for the two new pulsars discovered in NGC 6440, as obtained from fitting the observed ToAs with TEMPO2. The companion mass is calculated assuming a pulsar mass of 1.4 \msun. For both solutions, the time units are TDB, the adopted terrestrial time standard is TT(TAI) and the Solar System ephemeris used is JPL DE421 \citep{Folkner+2009}. Numbers in parentheses represent 1-$\sigma$ uncertainties in the last digit.}
    \centering
     \label{tab:timing_fitresults}
    \renewcommand{\arraystretch}{1.0}
     \begin{tabular}{l c c} 
     \hline
      \hline
Pulsar  &   NGC 6440G     & NGC6440H \\
\hline
R.A. (J2000) \dotfill   &  17:48:52.6460(4)  & 17:48:53.1995(1)  \\
DEC. (J2000) \dotfill   & $-$20:21:40.63(1)  & $-$20:21:35.31(5) \\
Proper Motion in $\alpha$, $\mu_\alpha$ (mas\,yr$^{-1}$) \dotfill &  1.8(1.2) &   $-$0.9(3) \\
Proper Motion in $\delta$, $\mu_\delta$ (mas\,yr$^{-1}$) \dotfill &   51(32) &  $-$11(8) \\
Spin Frequency, $f$ (s$^{-1}$)  \dotfill 	& 191.742146888623(1) &  351.06360053871(4)\\
1st Spin Frequency derivative, $\dot{f}$ (Hz\,s$^{-1}$) \dotfill & 5.8649(3)$\times 10^{-15}$ & $-$2.3413(1)$\times 10^{-14}$ \\
2nd Spin Frequency derivative, $\ddot{f}$ (Hz\,s$^{-2}$) \dotfill & -- & $-$2.04(2)$\times 10^{-25}$ \\
3rd Spin Frequency derivative, $\ddot{f}$ (Hz\,s$^{-3}$) \dotfill & -- & $-$1.8(9)$\times 10^{-34}$ \\
Reference Epoch (MJD)  \dotfill & 56668.558 & 56565.826\\
Start of Timing Data (MJD) \dotfill & 54050.705 & 54010.821\\
End of Timing Data (MJD) \dotfill 	& 59325.504 & 59184.670 \\
Dispersion Measure (pc\,cm$^{-3}$) \dotfill	& 219.601(2) & 222.584(7)\\
Number of ToAs \dotfill     & 111 & 111 \\
Weighted rms residual ($\mu$s) \dotfill  &  68.9 & 17.8 \\
$S_{1300}$ (mJy) \dotfill & 0.51 &  0.060\\
$L_{1300}$ (mJy kpc$^2$) \dotfill  & 3.52 & 4.15 \\
\hline
\multicolumn{3}{c}{Binary Parameters}  \\
\hline
Binary Model \dotfill   & -- & BTX \\
Orbital Period, $P_{\rm b}$ (days) \dotfill &  -- & 0.360787526(5) \\
Projected Semi-major Axis, $x_{\rm {p}}$ (lt-s)  \dotfill &  -- & 0.025324(3) \\
Epoch of Ascending Node, $T_\textrm{asc}$ (MJD) \dotfill &  -- & 58702.909365(1) \\
Orbital Frequency (Hz) \dotfill  & -- & 3.208002860(1)$\times 10^{-5}$ \\
\hline
\multicolumn{3}{c}{Derived Parameters}  \\
\hline
Spin Period, $P$ (ms)   \dotfill & 5.2153374530685(4) & 2.8484867085778(3) \\ 
1st Spin Period derivative, $\dot{P}$ (s\,s$^{-1}$)  \dotfill & --1.59523(7)$\times 10^{-19}$ & 1.89976(9)$\times 10^{-19}$\\
Mass Function, $f(M_{\rm p})$ (\msun)   \dotfill &  --  &  $1.339(4)\times 10^{-7}$  \\
Minimum companion mass, $M_{\rm c, min}$ (\msun) \dotfill &  -- & 0.0063\\
Median companion mass, $M_{\rm c, med}$ (\msun)  \dotfill &   -- & 0.0072 \\
Surface Magnetic Field\tnote{${\dagger}$}, $B_0$, (G)  \dotfill &   -- & $7.4\times 10^8$\\
Characteristic Age\tnote{${\dagger}$}, $\tau_{\rm c}$ (Gyr) \dotfill &   --  & 2.4 \\
    \hline
      \hline
     \end{tabular}
\begin{tablenotes}
      \small
      \item[${\dagger}$] The frequency derivatives could be affected by the cluster potential and therefore affect the values of  $B_0$ and $\tau_{\rm c}$.
\end{tablenotes}
  \end{threeparttable}
\end{table*}

\subsection{NGC 6440H}
\label{s: Timing_H}

To build the timing solution for NGC 6440H we first obtained two ToAs from each of the MeerTime observations where the pulsar was detected. The ToAs were then fitted with a pulsar model containing spin (frequency and first derivative) and orbital parameters ($x_{\rm p}$, $P_{\rm b}$ and $T_0$). The initial binary parameters were derived using \texttt{fit\_circular\_orbit.py} as described above. 

\begin{figure}
\centering
	\includegraphics[width=\columnwidth]{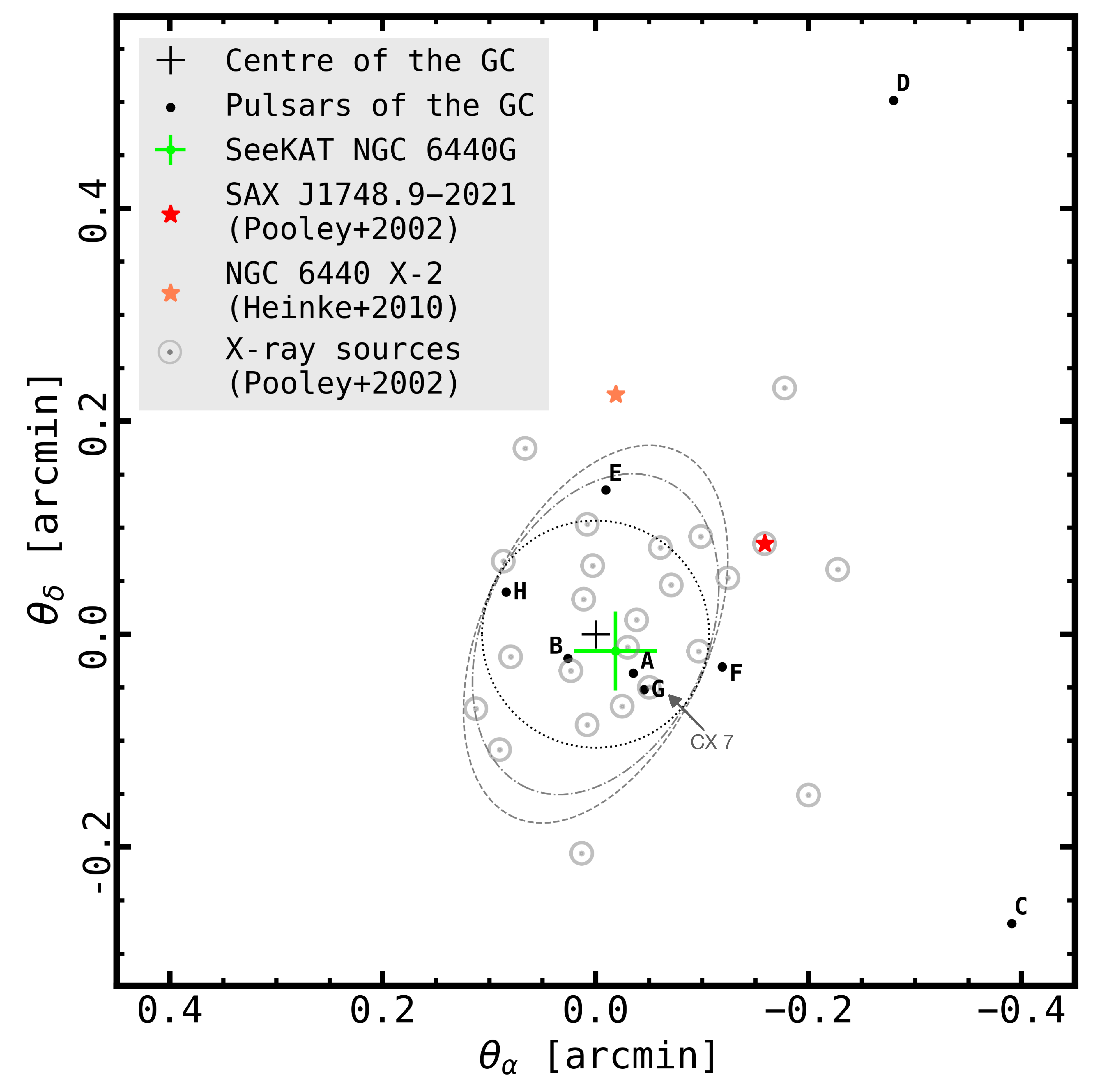}
    \caption{Positions of all the known pulsars in NGC 6440, plotted with respect to the centre of the GC. The localisation of NGC 6440G is highlighted in green, the size of the symbol represent the 2$\sigma$ uncertainty. The angular core radius ($r_{\rm c} = 0.10$\,arcmin) is indicated by the dashed circle. We show the size of the beam for 63 (inner ellipse) and 41 (outer ellipse) antennas at their 50 per cent power level. The two transient LMXBs known in the cluster are indicated with stars.}
        \label{fig:NGC6440_psrpositions}
\end{figure}

We could fold and detect the pulsar in a number of archival GBT observations of the cluster using the initial orbital model obtained with the MeerTime observations. We then used those GBT detections to generate the ToAs that were later used to obtain the long-term timing solution. For this, we first made use of the so-called ``jumps''\footnote{An arbitrary phase jump needed to take into account possible delays between the two different instruments.} between different epochs. This produced more refined orbital parameters, but we still needed to remove as many arbitrary jumps as possible by trying to estimate the exact number of rotations between the ToAs. We obtained a full timing solution after a few iterations of the same procedure. The minimum, median and maximum companion mass obtained from the \texttt{TEMPO2} fit were: 0.0063, 0.0072 and 0.0144\,\msun\ respectively, indicating that the companion is one of the lightest known. However, there is no evidence for eclipses in any of the observations, indicating that this pulsar is likely a non-eclipsing Black Widow. We were able to measure the pulsar astrometric and kinematic parameters thanks to the long timing baseline. NGC 6440H is located at R.A.=$17\h48\m$53$\fs$1995(1), Dec.=$-20\degr21\arcmin35\farcs31(5)$ which places it 0.09\,arcmin west of the cluster centre as shown in Figure \ref{fig:NGC6440_psrpositions}, which is near the edge of the core radius. The influence of the cluster on this pulsar appears to be small as the frequency derivative is about the right magnitude for an MSP (see Table \ref{tab:timing_fitresults}). The proper motion is loosely constrained, with measured values of $\mu_\alpha=-0.9\pm0.3$\,\pmunit~and $\mu_\delta=-11 \pm8$\,\pmunit, these values are consistent within errors with the values reported in \citet{Vitral2021}. Fitting for the orbital period derivative, $\dot{P}_{\rm b}$, we found that the uncertainty is greater than the fitted value. We also tried fitting higher orbital period derivatives but no significant values were obtained.

\subsection{NGC 6440G}

GBT archival data from both the GUPPI and Spigot backends from both the L-band and S-band receivers were also folded using the initial timing solution for NGC 6440G obtained with the MeerTime data. We could not phase-connect the GBT and MeerKAT data at this point. The weak nature of the pulsar meant that an accurate localisation was therefore needed. We then localised the pulsar using SeeKAT as described below. This allowed us to obtain a good enough ephemeris to fold the GBT data. After that, we obtained a timing solution of $\sim 14$ years using the combined GBT and MeerKAT data set (see Figure \ref{fig:NGC6440G_timingsol_MKT}) which resulted in a precise pulsar position.


\subsubsection{Localisation}
\label{ss:Localisation}

To obtain an accurate position for the source, we used a pipeline that applies the Tied Array Beam Localisation (TABLo) method, known as the SeeKAT multibeam localiser\footnote{\url{https://github.com/BezuidenhoutMC/SeeKAT}} (Bezuidenhout et al. in prep.).
This pipeline uses software called Mosaic\footnote{\url{https://gitlab.mpifr-bonn.mpg.de/wchen/Beamforming}} \citep{Chen+2021} to simulate the coherent beam tiling pattern and the Point Spread Function (PSF) of the telescope depending on the date, time and frequency of the observation, the location where the telescopes are pointing, the number of antennas used for the beamforming and the overlap fraction for placing the beams. SeeKAT uses the PSF, the coordinates (Right Ascension and Declination) and the S/N of the beams in which detections were made to determine the most likely position of the source. 

\begin{figure}
\centering
	\includegraphics[width=\columnwidth]{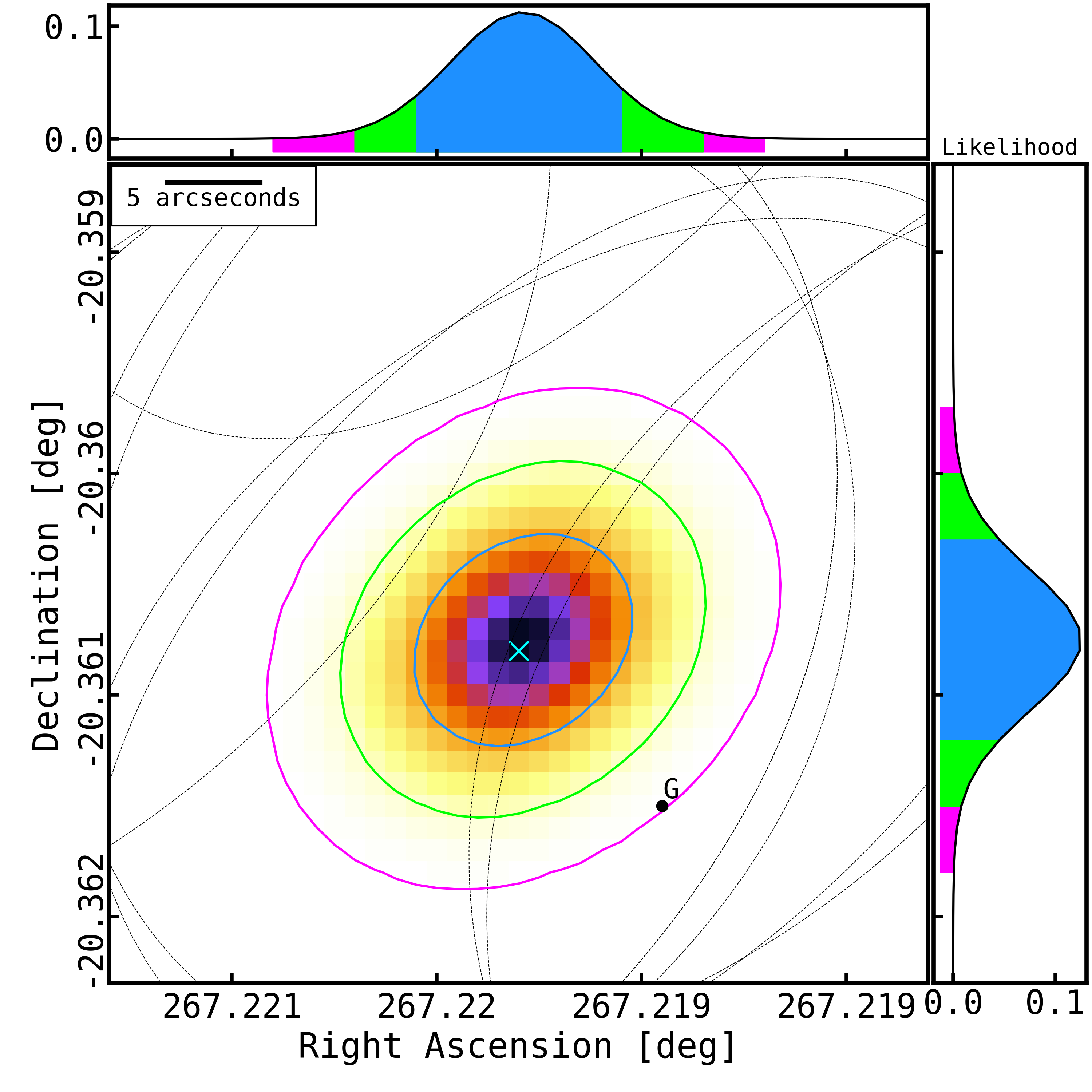}
    \caption{The localisation region of NGC 6440G obtained with SeeKAT. The 1-$\sigma$ error region is shown in blue, and the maximum likelihood position is highlighted by the cyan cross. The 2-$\sigma$ and 3-$\sigma$ error regions are shown by the green and magenta contours, respectively. Its timing position is also shown as a dot with the letter ``G''.}
    \label{fig:SeeKAT_multiepoch}
\end{figure}

 In the first follow up TRAPUM observations we pointed beam number 006 (of the 288 beams in total) towards the coordinates of the MeerTime beam (which was pointed at NGC 6440B; R.A. = $17\h48\m52\fs95$ and Dec. = $-20\degr21\arcmin38\farcs90$). However, the pulsar was detected with the highest S/N in TRAPUM beam number 008, with coordinates R.A.$=18\h23\m41\fs150$ and Dec.$=-30\degr21\arcmin38\farcs50$, which is nearer the GC centre. To get a more accurate localisation of the pulsar we used SeeKAT with both the long 2020 November observation and the 2020 December observation (see Table \ref{t:obs_summary}). These long observations used antennas outside of the core of the telescope and thus resulted in high-spatial resolution and good S/N. The source was detected in 4 and 7 beams in each of the observations respectively, and the beams overlapped at 70\% of the maximum sensitivity. Combining all the S/Ns from all the beams with detections from both epochs, the maximum likelihood localisation SeeKAT finds is R.A. $= 17\h48\m52\fs76  \pm 0.17$\,s, and Dec. $= -20\degr21\arcmin38\farcs45 \pm 2\farcs3$, as shown in Figure \ref{fig:SeeKAT_multiepoch}. The reported errors are obtained from the 2$\sigma$ region, shown in lime in  Figure \ref{fig:SeeKAT_multiepoch}, and the maximum likelihood position is highlighted with the cyan cross. This position places the pulsar closer to the centre of the GC ($\theta_{\rm c} = 0.02$\,arcmin). The long-term timing solution gives a position of $= 17\h48\m52\fs6460(4)$, and Dec. $= -20\degr21\arcmin40\farcs63(1)$ which is consistent with the SeeKAT results but places the pulsar a bit further from the cluster centre, at $\theta_{\rm c} = 0.07$\,arcmin.

\subsubsection{Proper Motion}

The $\sim$\,14--year timing baseline provided by the GBT data allowed low-precision estimates of the proper motion of NGC 6440G, $\mu_{\alpha} = 1.8(1.2)$\,mas\,yr$^{-1}$ and $\mu_{\delta} = 51(32)$\,mas\,yr$^{-1}$. By comparison, the proper motion of NGC 6440 has been estimated using Bayesian statistics by \citealt{Vitral2021} as $\mu_{\alpha} = -1.18(2)$\,mas\,yr$^{-1}$ and $\mu_{\delta} = -3.97(2)$\,mas\,yr$^{-1}$, for a total proper motion of 4.14(2)\,mas\,yr$^{-1}$. Therefore, the proper motion of NGC 6440G is unlikely to be real since the source is located within the core of the GC and such a large proper motion suggests it would have left the cluster within about 130\,years. One possibility is that it recently underwent some sort of dynamical interaction, but it is unlikely given the relative time scale.

\subsubsection{Period Derivative}

The intrinsic period derivative of pulsars in GCs can be contaminated primarily by the acceleration of the pulsar in the gravitational field of the cluster modifying the \Pdotobs. In order to account for the effects that contribute to the observed period derivative \Pdotobs\ we can use the following equation:

\begin{equation}
\label{eq:Pdot}
\left(\dfrac{\dot{P}}{P}\right)_{\rm obs} = \left(\dfrac{\dot{P}}{P}\right)_{\rm int} + \dfrac{a_{\rm GC} + a_{\rm G} + a_{\rm z} + a_{\rm PM}}{c},
\end{equation}

\noindent where \Pdotint\ is the pulsar's intrinsic period derivative. The various types of acceleration that contribute are: the line-of-sight component of the acceleration of the pulsar caused by the gravitational field of the cluster $a_{\rm GC}$, the radial acceleration due to the Galactic potential $a_{\rm G}$, the acceleration towards the Galactic plane $a_{\rm z} / c$, and the centrifugal acceleration due to the transverse Doppler effect \citep{Shklovskii1970} associated with the proper motion of the pulsar $\mu$ ($a_{\rm PM}$ =  $\mu^2 D$, where D is the GC’s distance from the Earth). The radial acceleration due to the Galactic potential is calculated using equation 2.3 in \citet{Phinney1992}, $a_{\rm G}/ c$ $\approx$ --1.99 $\times$ $10^{-18}$\,s$^{-1}$. Then, using equation 4 from \citet{Nice+Taylor1995}, we derive the acceleration towards the Galactic plane, $a_{\rm z} / c$  $\approx$ 1.62 $\times$ $10^{-19}$\,s$^{-1}$. We estimate $a_{\rm PM} / c$ $\approx$ 3.44 $\times$ $10^{-19}$\,s$^{-1}$ using the reported proper motion of NGC 6440 in \citet{Vitral2021} and the cluster distance \citep[see also][]{Lazaridis+2009}. The measured value of $(\dot{P}/P)_{\rm obs} = -1.49 \times 10^{-16}$\,s$^{-1}$, is much larger than the Galactic and Shklovskii terms and so the dominant contribution must be due to the cluster. 

In order to estimate the contribution from the GC potential, we consider a spherically symmetric cluster 

\begin{equation}
\label{eq:cluster_acc}
    \frac{a_{\rm GC}}{c} = -\frac{l}{c}\frac{GM(<r_{\rm psr})}{r^3_{\rm psr}},
\end{equation}

\noindent where $G$ is the gravitational constant, $r_{\rm psr} = \sqrt{R^2_{\bot} + l^2}$, is the distance from the pulsar to the centre of the cluster, $R_{\bot}$ is the projected distance on the sky plane between the cluster centre and the pulsar and $l$ is the distance between the pulsar and the plane of the sky passing through the cluster centre. The mass within the radius of the pulsar is given by 

\begin{equation}
\label{eq:mass_whithin}
    M(<r_{\rm psr}) = \int^{r_{\rm psr}}_0 4 \pi R^2 \rho (R) \dev R,
\end{equation}

\noindent where the volume mass density distribution is 

\begin{equation}
    \rho (R) = - \frac{1}{\pi} \int^{r_{\rm t}}_R \frac{\dev \Sigma(r)}{\dev r} \frac{1}{\sqrt{r^2 - R^2}} \dev r.
\end{equation}

\noindent We use a two-parameter King model \citep{King1962} to describe the surface mass density 

\begin{equation}
\label{eq:sup_massden}
    \Sigma (r) = \Sigma_0 \frac{1}{1 + \left(r/r_{\rm c}\right)^2},
\end{equation}

\noindent in this case $r_{\rm c}=0.26$\,pc. We can then use the cluster's total mass $M_{\rm tot}$ to make an estimate of the central surface mass density $\Sigma_0$ given by

\begin{equation}
\label{eq:sigma0}
    \Sigma_0 = \frac{M_{\rm tot}}{\int^{r_{\rm t}}_0 \frac{2\pi r\dev r}{1 + \left(r/r_{\rm c}\right)^2 }}.
\end{equation}

\noindent Equations \ref{eq:cluster_acc} to \ref{eq:sigma0} were developed in more detail in the appendix of \citet{Freire+2005}.

\begin{figure}
\centering
	\includegraphics[width=\columnwidth]{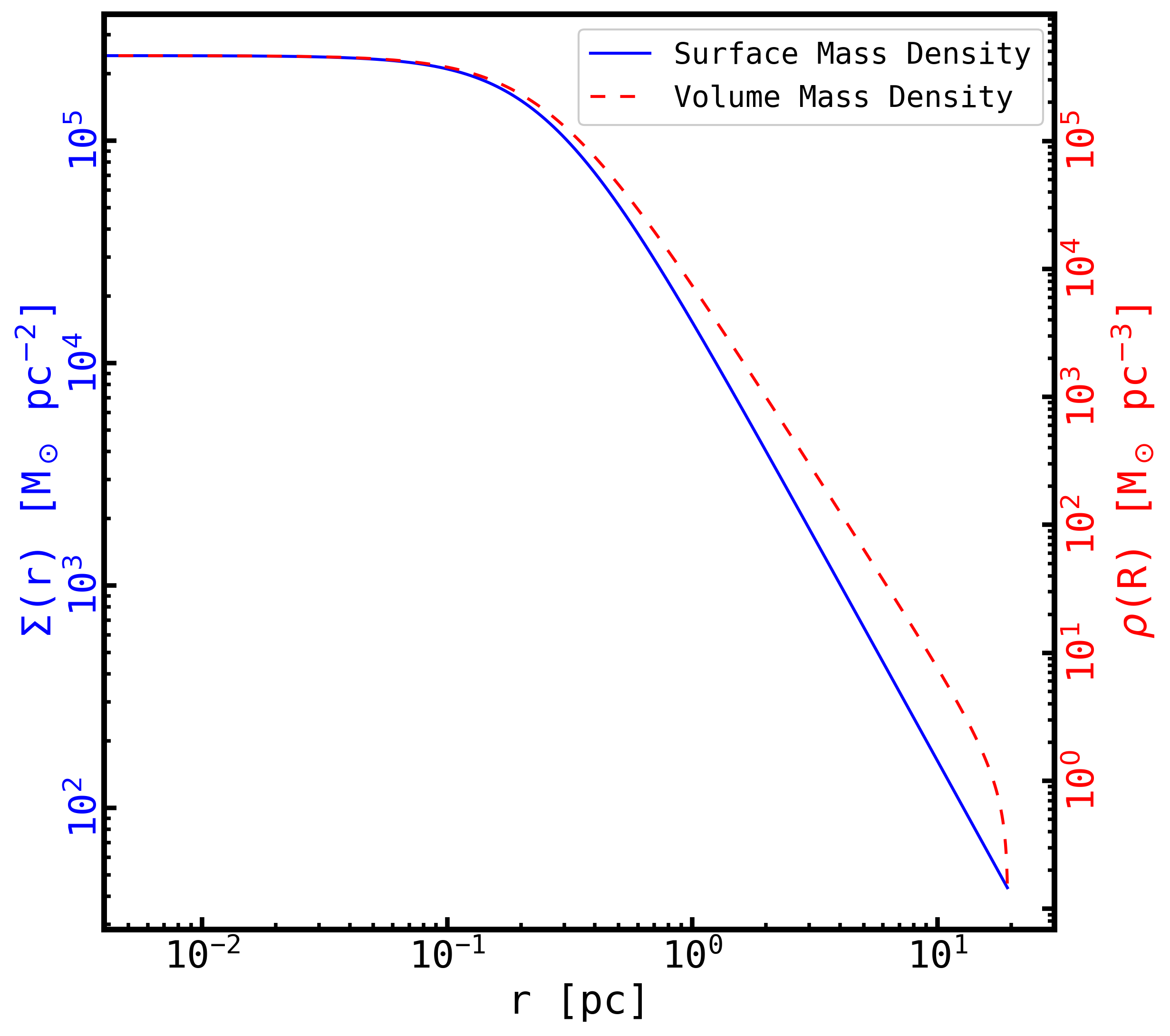}
    \caption{The surface mass density $\Sigma$ as a function of the distance to the cluster centre $r$ is shown as a blue solid line and enumerated on the left-hand side of the y-axis. The red dashed line shows the variation of the volume mass density distribution as a function of the distance $r$, and uses the right-hand side of the y-axis.}
        \label{fig:vol_massden_NGC6440}
\end{figure}

The values $r_{\rm c} = 0.26$\,pc and $r_{\rm t} = 19.4$\,pc were obtained from the most recent measurement of the basic parameters for NGC 6440 by \citet{Pallanca+2021} and $M_{\rm tot} = 4.42$ $\times$ $10^5$\,\msun\ from \citet{Baumgardt+Hilker2018}. The resulting $\Sigma_0$ $\approx$ $2.41$ $\times$ $10^5$\,\msun pc$^{-2}$ is then substituted to Equation~\ref{eq:sup_massden} to give us the surface mass density. The resulting variation of the volume mass density $\rho (R)$ as a function of the distance $r$ to the centre of the cluster is shown in Figure~\ref{fig:vol_massden_NGC6440}.

\begin{figure}
\centering
	\includegraphics[width=\columnwidth]{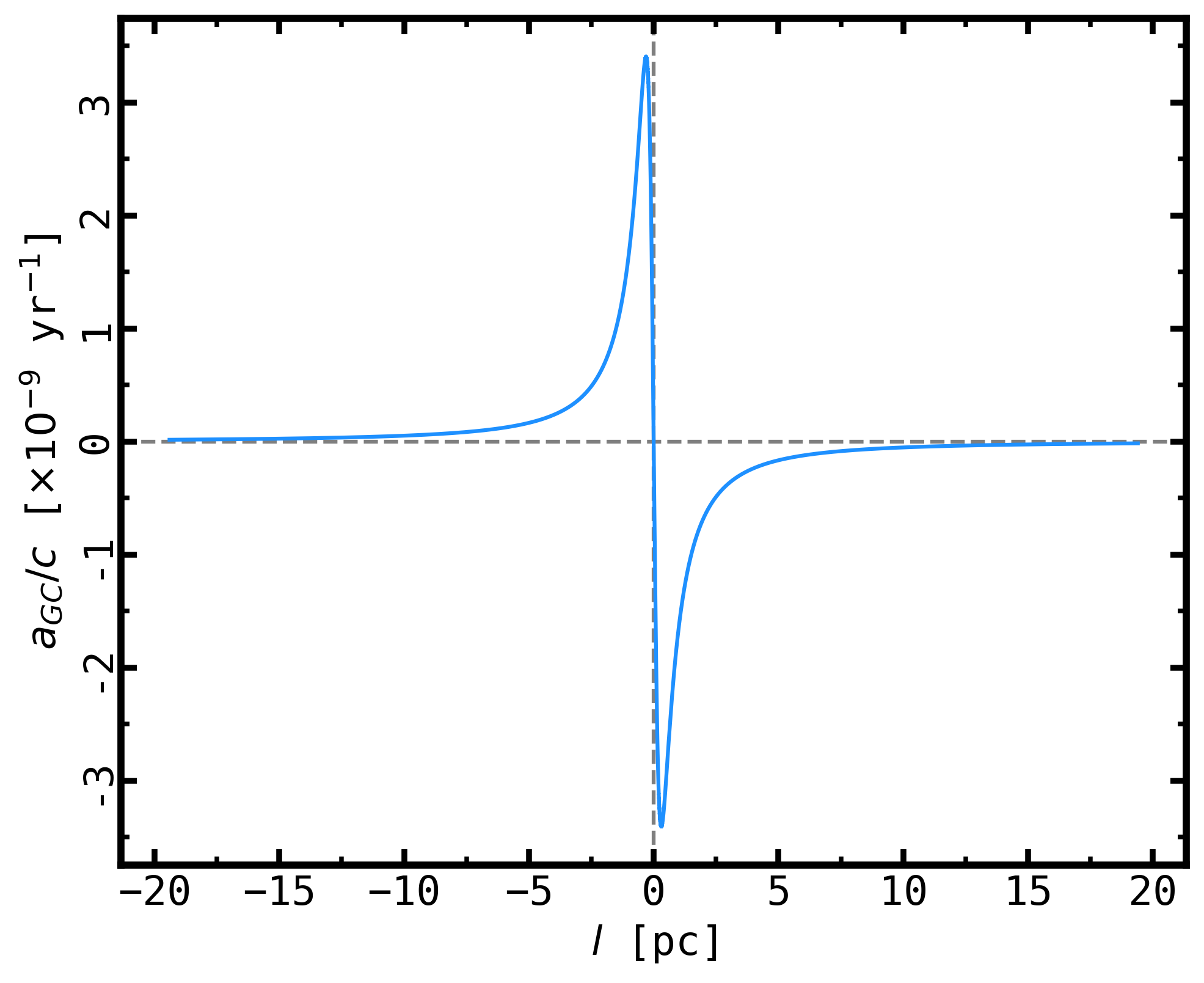}
    \caption{The line of sight acceleration due to the potential of the cluster as a function of the distance between the pulsar and the plane of the sky passing through the cluster centre $l$ for a measured $R_{\bot}$.}
        \label{fig:cluster_accel}
\end{figure}

We can use the volume mass density to calculate the mass within the radius of the pulsar (Equation~\ref{eq:mass_whithin}). However, since $r_{\rm psr}$ can not be directly measured, we used $R_{\bot} = 0.16$\,pc (see Table \ref{tab:pulsar_offsets}). We estimate the mass within the radius of NGC 6440G to be $M(< r_{\rm psr})$ $\approx$ $6.6$ $\times$ $10^3$\,\msun. We use this value to determine $a_{\rm GC}/c$ as a function of $l$ (see Figure \ref{fig:cluster_accel}).

Since we do not know the exact value of $l$, we can calculate the maximum acceleration expected near the GC centre ($R_{\bot} < 2 r_{\rm c}$, within $\sim 50$\%) by using equation 2.7 from \citet{Phinney1992}. Since we do not have an accurate value of the line-of-sight velocity dispersion at the pulsar position $v^2_l(R_{\bot})$, we use the value of the velocity dispersion of the cluster core $v_{\rm z}(0)$ assuming that the dispersion does not change significantly at our small distance from the core. We take the value $v_{\rm z}(0) = 13.01$\,km\,s$^{-1}$ from \citet{Webbink1985}, and as a result we estimate $\abs{a_{\rm GC}}/ c = 8.94$ $\times$ 10$^{-17}$\,s$^{-1}$. This value is of the same order of magnitude to the observed value of  $\abs{\dot{P}/P} = 3.05 \times 10^{-17}$\,s$^{-1}$, which agrees with our assertion that the observed period derivative is dominated by acceleration in the cluster.
The negative value of the $(\dot{P}/P)_{\rm obs}$ indicates that it is on the back side of the cluster, but close to the centre (see Figure~\ref{fig:cluster_accel}).
For comparison we calculate the absolute upper limit on the acceleration $a_{\rm{GC,max}}$ and the maximum intrinsic period derivative \Pdotint\ for all the pulsars studied in this work (see Table \ref{tab:derivaded_params}).

\begin{table*}
\caption{Limits for derived parameters of the pulsars in NGC 6440 studied in this paper.}
\label{tab:derivaded_params}
\footnotesize
\centering
\renewcommand{\arraystretch}{1.0}
\vskip 0.1cm
\begin{tabular}{lccccccc}
\hline
\hline
Pulsar   & $P$ & \Pdotobs & 
$a_{\rm p,max}$ & $a_{\rm GC,max}$ &  $\dot{P}_{\rm int,max}$   & $B_{\rm max}$  & $\tau_{\rm c,min}$   \\
 & (ms) & (10$^{-18}$) & (10$^{-9}$ m s$^{-2}$) &  (10$^{-9}$ m s$^{-2}$) & (10$^{-18}$)  & (10$^9$ G) & (Gyr)  \\
\hline
C   &  6.2269328180(7) &  -0.062(2) & -2.89 &  6.99  & 0.08 &  0.74  & 1.15   \\ 
D   &  13.495818491(3) & 0.558678(16) & 13.03 & 5.83 & 0.85 &  3.43  & 0.25   \\
G    & 5.2153374530685(4) & -1.59523(7) & -9.17 &  26.80 & 0.31 & 1.28 & 0.27   \\
H    &  2.8484867085591(7) & 1.89958(3) & 20.00 &  24.15  & 0.42 &  1.10  &  0.10  \\
\hline
\hline
\end{tabular}
\end{table*}

\subsection{Possible counterparts and location in the cluster}

Figure \ref{fig:NGC6440_psrpositions} shows the positions of all the known pulsars in the cluster relative to its centre. We also show the positions of the known X-ray sources in this GC as grey dots, and the grey circles show their positional uncertainty (from \citealt{Pooley+2002}). Using our best position we find that NGC 6440G is potentially associated with source CX 7 in \citet{Pooley+2002}. The positional offset is 0.32\,arcsec and the uncertainty is $1.3$\,arcsec. CX 7 has an X-ray luminosity of $L_{\rm X} =$ 2.0 $\times$ $10^{32}$ ($\pm$ 21\%) $\ergs$. Using the observed relation  $L_{\rm X} \approx 10^{-3}L_{\rm sd}$, where $L_{\rm sd} \equiv I(2\pi)^2\dot{P}/P^3$ is the spin-down luminosity of the pulsar \citep{Verbunt+1996}, this implies a $L_{\rm sd} \sim 10^{35}$\,$\ergs$. Then if the sources are associated, the period derivative of the pulsar should be of the order $\sim 10^{-19}$\,s\,s$^{-1}$.

In Table \ref{tab:pulsar_offsets}, we report the angular offsets of all the pulsar positions relative to the centre of the GC. The projected distances ($R_\bot$) were calculated using the most recent GC parameters from \citet{Pallanca+2021}. The positions of the known pulsars were obtained from \citet{Freire+2008}. We note that the two pulsars discovered in this work are located within the core radius of the GC.

\begin{table}
\setlength\tabcolsep{8pt}
  \begin{threeparttable}
    \caption{Pulsar offsets from the centre of NGC 6440.}
     \label{tab:pulsar_offsets}
    \renewcommand{\arraystretch}{1.0}
     \begin{tabular}{lccccc}
     \hline
      \hline
 & & & \multicolumn{2}{c}{$\theta_{\bot}$} & \\
 & $\theta_{\alpha}$\tnote{*} & $\theta_{\delta}$\tnote{*} &  \multicolumn{2}{c}{\rule{2cm}{0.2pt}} & $r_{\bot}$   \\
Pulsar & (arcmin) & (arcmin) & (arcmin) &  ($\theta_{\rm c}$) & (pc)  \\        
        \midrule
A    &  -0.0354  & -0.0366 &  0.05  &  0.48  &  0.12 \\
B    &  0.0262  & -0.0226 &  0.03  &  0.33  &  0.08  \\
C    &  -0.3907  &  -0.2718 &  0.48  &  4.46  &  1.15  \\ 
D    &  -0.2799  & 0.5015  &  0.57  &  5.38  &   1.39  \\
E    &  -0.0093  &  0.1353  &  0.14  & 1.27  &  0.33  \\
F    &  -0.1188  &  -0.0305  &  0.12  &  1.15  &  0.30  \\
G    &  -0.0454  & -0.0522  &  0.07  &  0.65  &  0.17  \\
H    &  0.0842  & 0.0395  &  0.09  &  0.87  &  0.22  \\  
    \hline
      \hline
     \end{tabular}
    \begin{tablenotes}
      \small
      \item[*] The uncertainties are much smaller than the uncertainty of the GC's centre, assumed to be exactly where indicated in Figure \ref{fig:NGC6440_psrpositions}, except for the case of NGC 6440G. The 1-$\sigma$ error for the latter are $\theta_{\alpha} = 0.021$\,arcmin and $\theta_{\delta} = 0.019$\,arcmin.
    \end{tablenotes}
  \end{threeparttable}
\end{table}

\subsection{Flux Densities}
\label{s:flux_densities}

\begin{figure}
	\includegraphics[width=\columnwidth]{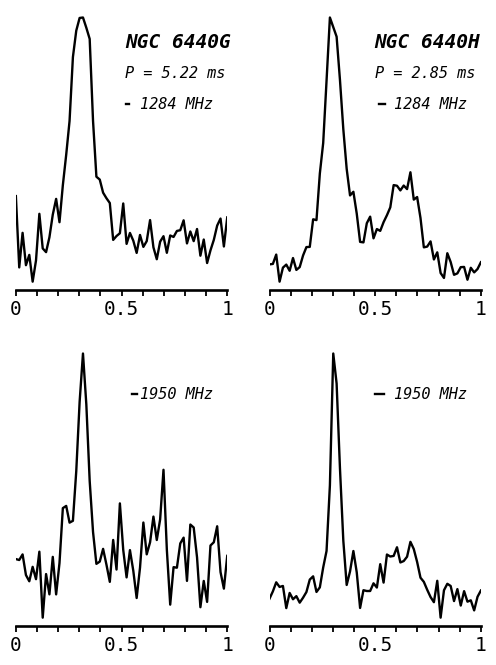}
    \caption{Integrated pulse profiles for the newly discovered pulsars presented here using both L-band (1284 MHz) and S-band (1950 MHz) data from MeerKAT and GBT, respectively. All profiles show one full rotation of the pulsar with 64 phase bins. Their spin periods are indicated. The width of the horizontal bar indicates the effective time resolution of the system, relative to each pulsar’s spin period.} 
    \label{fig:integrated_profiles}
\end{figure}

The flux densities at $\sim 1300$\,MHz ($S_{\rm 1300}$) of the new pulsars and their pseudo-luminosities ($L_{1300} \equiv S_{1300}D^2$) are calculated using the radiometer equation \citep{Dewey+1985} and using the cluster distance from \citet{Pallanca+2021}. To calculate the system equivalent flux density at 1300\,MHz, we used a system temperature $T_{\rm sys}$ = 26\,K, which includes the atmosphere plus the ground spillover temperature $T_{\rm atm+spill}$ ($\sim 4.5$\,K at 45\degr elevation\footnote{{\bf \url{https://skaafrica.atlassian.net/rest/servicedesk/knowledgebase/latest/articles/view/277315585} }}), the receiver temperature $T_{\rm rec}$ (18\,K), the cold sky temperature $T_{\rm sky}$ ($\sim 3.5$\,K at 1.3\,GHz), and the gain of the telescope $G = 2.59$\,K\,Jy$^{-1}$ (for the MeerKAT array observations using 59 antennas). To account for the various sensitivity losses due to signal processing and digitisation, we assumed a correction factor of 1.1. 
NGC 6440G has an estimated mean flux density $S_{\rm 1284} = 0.012$\,mJy and a pseudoluminosity $L_{1284} = 0.83$\,mJy\,kpc$^2$, whereas for NGC 6440H, $S_{\rm 1284} =$ 0.020\,mJy and $L_{1284} = 1.43$\,mJy\,kpc$^2$ (see Table~\ref{tab:timing_fitresults}). The results were determined using the integrated pulse profiles shown in the top panels from Figure \ref{fig:integrated_profiles}, obtained by summing together (without including weights) a total of 58\,hrs of observation for both NGC 6440G and NGC 6440H, equivalent to all MeerTime observations which resulted in a detection. 
For comparison, the 1950-MHz flux densities and their pseudo-luminosities were also obtained using the GBT data and the $T_{\rm sys}$, the gain of the telescope, the bandwidth and the correction factor specified in \citet{Freire+2008}. The average pulse profiles used in this case are shown in the bottom panels from Figure \ref{fig:integrated_profiles}, obtained by adding all observations made with Spigot, equivalent to 20\,hrs of observation for NGC 6440G. In the case of NGC 6440H, we summed together all the GUPPI observations dedispersed at 1950 MHz, summing a total of 40\,hrs. We did not include weights in any of the cases. The estimated mean flux density of NGC 6440G at 1950\,MHz $S_{\rm 1950} = 0.006$\,mJy and the pseudoluminosity $L_{1950} = 0.40$\,mJy\,kpc$^2$, whereas for NGC 6440H, $S_{\rm 1950} = 0.009$\,mJy and $L_{1950} = 0.60$\,mJy\,kpc$^2$.

\subsubsection{Profile Widths}

To measure the observed pulse width at 50\% of the peak intensity (W50) we again used the template we formed to perform the timing analysis. Using the concentration parameter from the best fit combination of von Mises functions one can determine the pulse width using the \texttt{fitvonMises} function from \texttt{PSRSALSA} \citep{Weltevrede2016}. To calculate the error on the width, we generated 1000 simulated noise profiles using the off-pulse noise statistics of the average profile and repeated the von Mises fitting, and determined the width in each case. The error was then estimated from the distribution of width values. The resultant widths are, at 1284\,MHz, for NGC 6440G W50 = (6.31 $\pm$ 0.36) $\times$ 10$^{-4}$\,s, while for NGC 6440H W50 = (2.85 $\pm$ 0.10) $\times$ 10$^{-4}$\,s, which corresponds to a duty cycle of 12\% and 10\%, respectively. Whereas at 1950\,MHz, for NGC 6440G W50 = (3.05 $\pm$ 0.76) $\times$ 10$^{-4}$\,s while for NGC 6440H W50 = (1.40 $\pm$ 0.04) $\times$ 10$^{-4}$\,s, which corresponds to a duty cycle of 6\% and 5\%, respectively.

\subsection{High-cadence data analysis for NGC 6440C and NGC 6440D}

 Previous works (e.g. \citealt{Freire+2008}, \citealt{Begin2006}) have presented interesting systematics with time that could be explained by the presence of unmodelled planetary companions (e.g. \citealt{Cordes1993}). The high cadence and high sensitivity MeerTime data provided us an opportunity to detect, or place tight constraints on the presence of planets around NGC 6440C. Similarly, these data and the wide bandwidth of the L-band receiver allowed us to study the eclipses of NGC 6440D in more detail. In this section we present the results of these analyses.

\subsubsection{NGC 6440C}

We know that any unmodelled pulsar companions, depending on their mass, may produce orbital effects that can be identified in the timing residuals. NGC 6440C is an apparently isolated MSP with a spin period of 6.23 ms.  With the aim of investigating the nature of the systematics in the timing behaviour, we carried out searches for planets using a code which is based on ENTERPRISE \footnote{\url{https://github.com/nanograv/enterprise/tree/master/enterprise}}\citep[Enhanced Numerical Toolbox Enabling a Robust PulsaR Inference SuitE,][]{enterprise}. It uses the Monte Carlo Markov Chains (MCMC) technique to fit for the parameters of planets like their period, mass, the white noise and the red noise, among others, using the timing solution of the pulsar \citep{Nitu+2022}. We tried different period and mass ranges that were obtained by looking for the maximum amplitude sine wave that can be hidden in the residuals and the minimum amplitude sine wave we might expect based on the ToA errors. These limits are used as input to the mass function equation:

\begin{equation}
    f(M_{\rm p}) = \dfrac{(M_{\rm c}\sin i)^3}{(M_{\rm p} + M_{\rm c})^2} = \dfrac{4\pi^2}{T_{\odot}} \dfrac{x^3}{P_{\rm b}^2},
\end{equation}

\noindent where the constant $T_{\odot} = GM_{\odot}/c^3 = 4.925\,\mu$s is used to express the masses in solar units. The red noise was not included in the fit.

We first use the ephemeris from the timing solution obtained using 1 TOA for each of the 26 MeerTime detections, resulting in a data span of 42\,days. We did not initially use the TRAPUM data since we first focused on the high-cadence data. For orbital periods ranging between 5 to 35\,days, and masses between $1\times10^{-3}$ to 10\,\mearth, our analysis does not find any evidence for planets. 

\begin{figure}
\centering
	\includegraphics[width=\columnwidth]{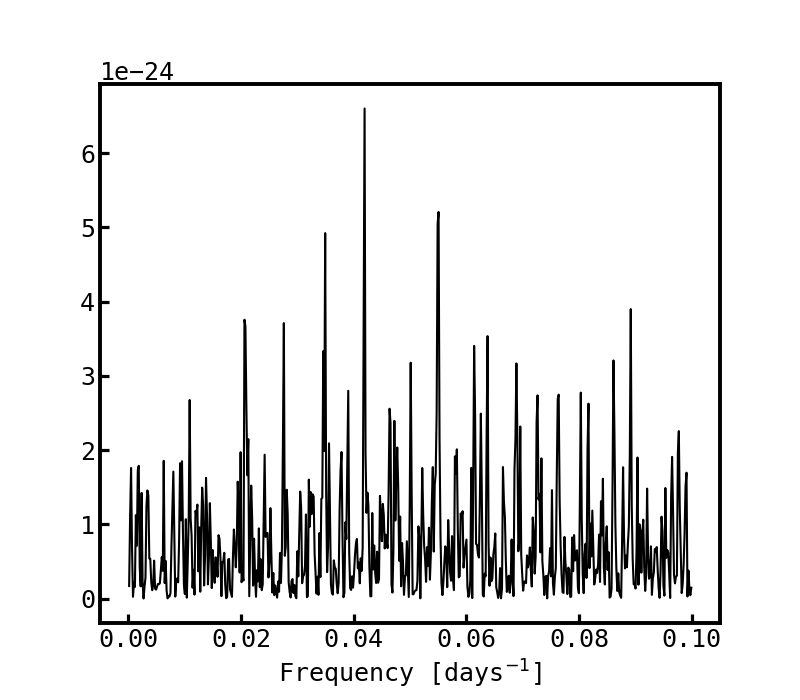}
    \caption{Generalised Lomb-Scargle calculated from the timing residuals from the best fit timing solution to the NGC 6440C 16-year data set, obtained using \texttt{cholSpectra} from \TEMPOTWO.}
        \label{fig:GLS_NGC6440C}
\end{figure}

We then expanded the data set by adding all the GBT and TRAPUM data available for this source. With this, we obtained a long-term timing solution over 16\,years. As an example of what is seen in the data, we calculated the Generalised Lomb-Scargle periodogram \citep{Lomb1976, Scargle1982} using \texttt{cholSpectra} from \TEMPOTWO. From the plot shown in Figure~\ref{fig:GLS_NGC6440C}, we see a peak at 23.84\,days but it is only $\sim 1.3 \sigma$; inspection of the residuals shows that there is some structure on those timescales, but the signal is clearly weak. We note that this is close to the value of 21.627\,days reported by \citet{Begin2006}.

\begin{figure}
\centering
	\includegraphics[width=\columnwidth]{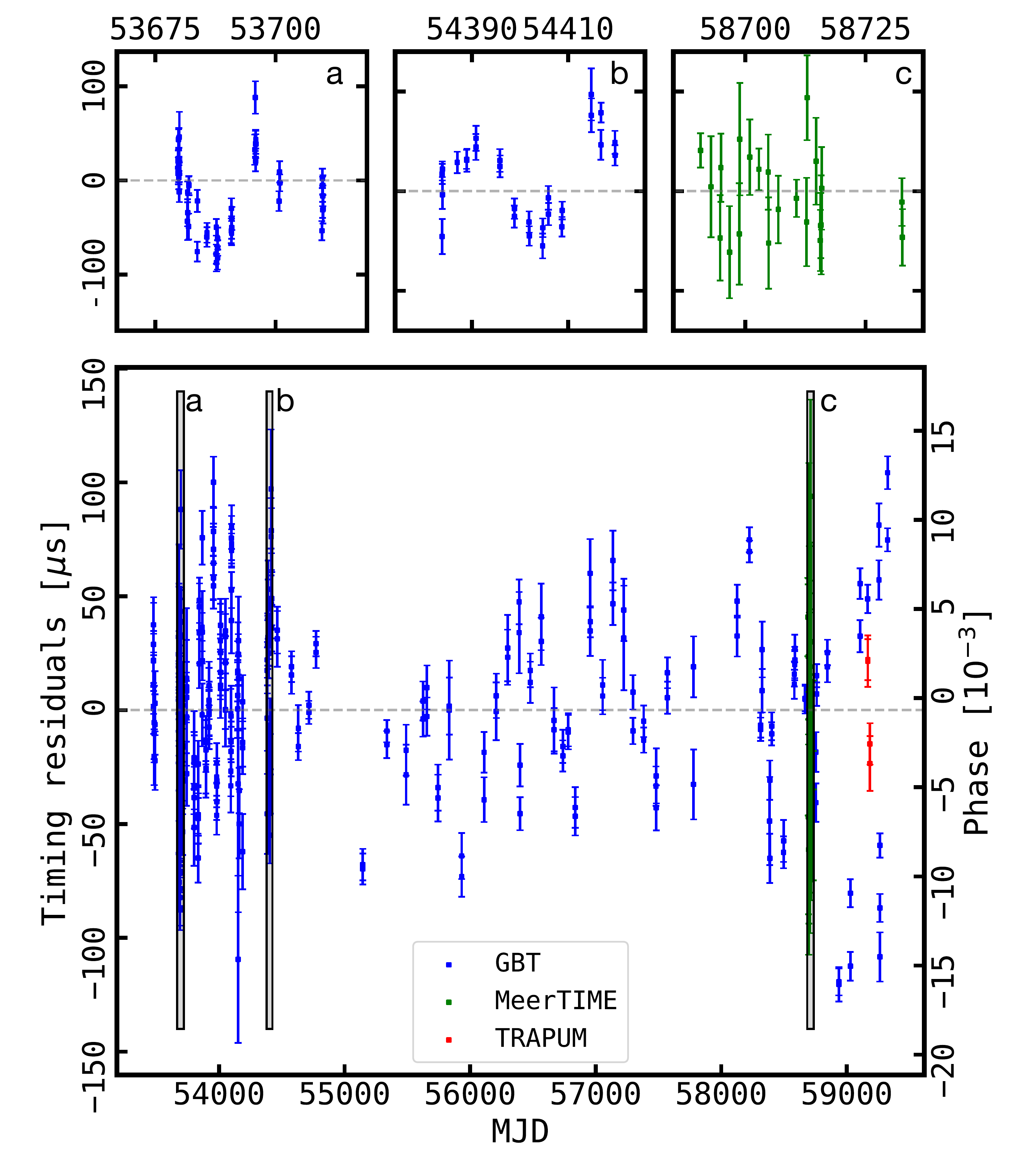}
    \caption{Timing residuals of NGC 6440C as a function of time for GBT and MeerKAT data. The top panels correspond to zoom-in portions of the rectangles labelled as a, b and c in the main panel.}
        \label{fig:zoom_timing_NGC6440C}
\end{figure}

We did the same planetary search using \texttt{ENTERPRISE} and the dynesty sampler fitting for a planet with period between 5 to 35\,days and mass between $1 \times 10^{-3}$ and 10\,\mearth. This fit revealed the possibility of two planets orbiting NGC 6440C, i.e. two different periodicities in the data, so more investigation was warranted. The GBT data were split into two segments of equal number of TOAs. The fits for the first half gave a period of $\sim 18.4$\,days, while the second half showed a period of $\sim 36.2$\,days, two periodicities that are not actually harmonically related. 
Figure~\ref{fig:zoom_timing_NGC6440C} shows the long-term timing residuals of NGC 6440C as function of time for GBT and MeerKAT data. If we only consider the infrequent observations over the total observation span, it appears that there is excess white noise in the data. However, the high cadence observations shown in the inset panels of Figure~\ref{fig:zoom_timing_NGC6440C} show that the noise is correlated on timescales of tens of days. Correlated noise in pulsars typically takes the form of a red noise process, but here we see no long timescale fluctuations. Given that correlated structures in three sessions of high cadence observations with the GBT (panels a and b in Figure \ref{fig:zoom_timing_NGC6440C}) and MeerKAT (panel c in Figure \ref{fig:zoom_timing_NGC6440C}) are not consistent with a purely periodic behaviour, we suggest that this pulsar exhibits some kind of band-limited noise, or a weekly quasi-periodic process. This might imply association with slow changes in the pulse profile, and hence correlated phase jitter, rather than fluctuations in the spin down rate typically associated with timing noise. We leave the study of the pulse shape for future work. We note that other pulsars in the cluster observed in the same programmes do not show this excess noise and hence it is unlikely to be related to instrumentation or an artefact of the observing cadence.

\subsubsection{NGC 6440D}
\begin{figure}
\centering
	\includegraphics[width=\columnwidth]{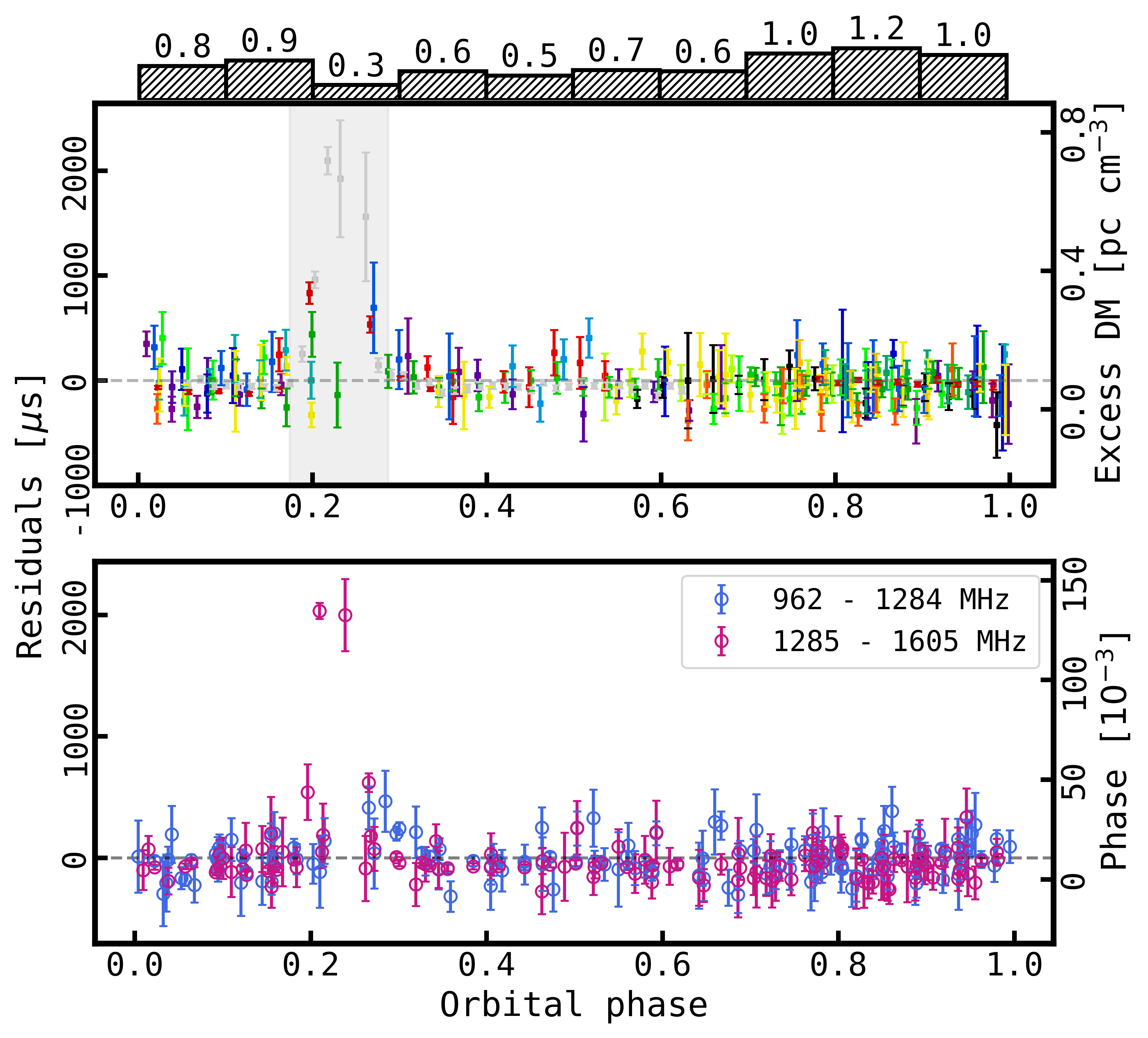}
    \caption{Top panel: Timing residuals and excess DM for the eclipsing binary pulsar NGC 6440D as a function of orbital phase using the frequency range 962-1605\,MHz. The eclipse region is highlighted with a grey coloured region. The different epochs are denoted by different colours. The histogram at the top shows a density of ToAs that is not uniform which is related to the timing of the observations. Bottom panel: Variation of timing residuals and phase with orbital phase measured simultaneously in two subbands: 962-1284\,MHz (blue circles) and 1285-1605\,MHz (pink circles).} 
    \label{fig:NGC6440D}
\end{figure}

Taking advantage of the sensitivity and wide band of MeerKAT, we study the eclipses of the Redback NGC 6440D in more detail than previous works \citep{Begin2006,Freire+2008}. Our data set consists of 18 epochs of observations at the central frequency of 1284\,MHz. The timing solution of these observations gave us an orbital period $P_{\rm b} =$ 0.286068574(2) days and a projected semi-major axis of $x = 0.39722(1)$\,lt-s. Delays of $\sim 1$\,ms in the times of arrival can be seen before, after, and sometimes through the eclipse. The maximum delay in timing residuals around eclipse transitions is $\sim 2$\,ms at $\sim 0.22$ in phase as visible in the top panel of Figure \ref{fig:NGC6440D}. The variation in the excess DM with orbital phase derived from the best-fit timing residuals is also shown in this Figure. The presence of an additional electron column density in the eclipse region implied by this dispersive delay, that is, the maximum added electron density near superior conjunction, is found to be $N_{e,{\rm max}} \geq$ 2.13 $\times$ 10$^{18}$\,cm$^{-2}$. We estimate the corresponding electron density in the eclipse region ($n_e \sim N_e/2a$, where $a = a_{\rm p} + a_{\rm c} \approx 2.1$\,\rsun, assuming an inclination $i = 90\deg $) as 8.0 $\times$ 10$^8$\,cm$^{-3}$. The eclipse ingress and egress transitions are spread over a range of orbital phases. The ingress transition starts at orbital phase $\sim 0.17$, while the eclipse egress ends at $\sim 0.29$, giving a total span of the eclipse of approximately 12\% of the orbit.

Using the mass function from timing $f(m_{\rm p},m_{\rm c}) = 8.2233(7)$ $\times$ 10$^{-4}$\,\msun\ and assuming a pulsar mass $m_{\rm p} = 1.4$\,\msun\ and inclination angle $i = 90^{\circ}$, we obtain a minimum companion mass $m_{\rm c} = 0.12$\,\msun. The separation of the binary components corresponds to an eclipsing region with physical size of R$_{\rm E}$ $\sim 0.8$\,\rsun, which is larger than the Roche lobe radius of the companion R$_{\rm L} \simeq 0.5$\,\rsun, indicating that the eclipsing material is not gravitationally bound to the companion and that the companion is losing mass \citep{Freire2005}. We also calculate the energy density of the pulsar wind at the companion distance, $U_{\rm E} = \dot{E} / 4\pi^2 ca^2$, and obtain $U_{\rm E} = 2.50$\,erg\,cm$^{-3}$ using the $\dot{P}$ from \cite{Freire+2008} since our $\dot{P}$ estimation would be biased due to the short-term solution.
In Figure \ref{fig:NGC6440D}, we notice that the number of ToAs is significantly less around superior conjunction compared to other orbital phases and an increase of ToAs is seen after inferior conjunction with a maximum density of ToAs at orbital phases 0.7 - 0.9. We attribute the excess at inferior conjunction to the way the observations were taken, as they were optimised for studying NGC 6440B, with $\sim 60$\% of the observations covering the orbital phases among 0.6 and 1.

In order to study the frequency dependence of the eclipse of NGC 6440D we divided the observed bandwidth (643\,MHz) into two sub-bands as seen in the bottom panel of Figure \ref{fig:NGC6440D}. The maximum delay in timing residuals around eclipse transitions is $\sim 2$\,ms at $\sim 0.22$ in phase, this is seen only at higher frequencies (1284-1605 MHz, pink colour in Figure \ref{fig:NGC6440D}). On the contrary, at lower frequencies (962-1283\,MHz, blue colour in Figure \ref{fig:NGC6440D}), we see a full eclipse. We also observe a larger eclipse duration at the lower frequency band ($\sim 1.2$ times longer for the 962-1283\,MHz band than in the 1284-1605\,MHz band). We note a longer egress in the 962-1283\,MHz band compared to the 1284-1605\,MHz band in which case the egress is very quick. The eclipse egress duration at lower frequencies is 23.61 $\pm$ 2.54\,minutes. Different ingress/egress duration that depend on the frequency have also been observed for other spider pulsars \citep[e.g.][]{Kudale+2020,Polzin+2020}. 

\section{Discussion}
\label{s:Discussion}

This paper has presented two new MSPs and an analysis of timing data for two of the known pulsars in NGC 6440. The following section will discuss the implications of the results obtained for each of the pulsars studied in this work and examine what the pulsars tell us about the cluster.

NGC 6440G was found to be an isolated pulsar, it has a positional overlap with the X-ray source CX 7 from \citet{Pooley+2002} which is proposed as a quiescent LMXB by the authors. The X-ray luminosity of the possible associated source implies a period derivative of the pulsar of the order of $\sim 10^{-19}$ s s$^{-1}$, a typical value for the MSP population. It is then consistent with the possibility of the association of the source CX 7 with NGC 6440G. However, since the \Pdotobs~ value we have is contaminated by the gravitational potential of the cluster, we can not confirm the association between the sources. If indeed CX 7 is a qLMXB then the two sources cannot be associated.

Many of the low-mass systems are either Black Widows or Redbacks and they exhibit eclipses. The NGC 6440H 1284-MHz ToAs are well described by a circular orbit and do not show eclipses around superior conjunction. Pulsars with similar characteristics (short orbital period, low-mass companion and no eclipses) can be found in globular clusters, e.g., 47 Tuc I, 47 Tuc P, M62 F \citep{Lynch+2012}, and NGC 6544A which also has a companion with planetary mass (M$_{\rm c}$ $\sim$ 10\,M$_{\rm J}$; \citealt{DAmico+2001}). The 90 per cent confidence upper limit on the companion mass for NGC 6440H was determined by assuming an inclination angle of 26\degr. With this, we find an upper limit on the companion mass of $\sim 14$\,M$_{\rm J}$. The nature of the companion could possibly be explained as a ultra-low-mass carbon white dwarf or a brown dwarf.

There are 57 known pulsars with ultra-light mass companions (M$_{\rm{c,min}}$ < 0.08\,\msun) and only 9 pulsars\footnote{This number does not include pulsar B1620--26 since the mass of planetary companion is not well known and it is not included in the ATNF catalogue.} with minimum companion masses smaller than 0.01\,\msun \citep{Manchester+2005}. The pulsars with the lightest companions are PSR J2322--2650, with a minimum companion mass of 0.0007\,\msun\ \citep[0.7\,M$_{\rm J}$;][]{2018MNRAS_Spiewak+}, and the ``diamond planet'' orbiting PSR J1719--1438 with  M$_{\rm {c, min}}= 0.0011$\,\msun\ \citep[1.2\,M$_{\rm J}$;][]{Bailes+2011}. The minimum companion mass of NGC 6440H makes it the third lowest among all known pulsars\footnote{\url{https://www.atnf.csiro.au/research/pulsar/psrcat/}} and the lowest minimum companion mass in a GC\footnote{See \url{http://www.naic.edu/~pfreire/GCpsr.html}} with M$_{\rm{c, min}} \sim 6.6$\,M$_{\rm J}$. B1620--26 is another pulsar with a planetary mass companion located in a GC, M4. It is in a triple system with one companion of a few M$_{\rm J}$ in a many year orbit which was most likely captured \citep[see e.g.][]{Thorsett+1999,Sigurdsson+2003, Sigurdsson+2008}. 

The fact that both NGC 6440G and NGC 6440H were missed in the GBT data was expected, since they are very faint even after folding with a precise ephemeris (see Section \ref{s:flux_densities}). The archival GBT data can recover the signal from the pulsars only after the orbits were first characterised in the MeerKAT observations, decreasing acutely the number of trials necessary for a detection. With this, a suitable set of ToAs can be determined. These can be then used to derive a long-term timing solution with precise timing parameters.

The structure in the timing residuals of NGC 6440C are not well modelled as a standard power-law-like noise process but are also not modelled by a planet, or planets. The quasi-periodic nature of the variations argues against them being due to random dynamical interactions with passing stars. So the origin of the unusual variations in arrival times is unclear and probably requires a dedicated long term and high-cadence, high-sensitivity observing campaign to better constrain the properties and hence the origin. 

The detection of NGC 6440D at higher frequencies during the eclipse phase indicates an orbital inclination that is likely less than 90$^{\circ}$ and it implies a highly variable concentration of eclipsing material. This has also been seen in other similar systems. Moreover, the longer egress duration at lower frequencies suggest that the companion’s orbital motion causes its wind to be swept back (\citealt{Fruchter+1990}; \citealt{Stappers+2001}) leaving a cometary-like tail of material \citep{Main+2018} causing the eclipses to be extended. At higher frequencies typical pulse smearing effects like scattering or dispersion would be less prominent. Since the eclipse is not always total (see Figure \ref{fig:NGC6440D}) the eclipse could be caused by grazing incidence of the pulsar beam on a wind from the companion.

Comparing the values of the eclipse radius, Roche lobe radius, orbital period P$_{\rm b}$, energy density of pulsar wind at the companion distance $U_{\rm E}$ and $N_{\rm e,max}$ for NGC 6440D with those of other eclipsing Redback pulsars from the Galaxy, we find that this MSP is similar in most respects (see \citealt{Polzin+2020}). This pulsar is far away from the cluster centre and its period derivative is not highly affected by the gravitational potential of the cluster.

Besides the 8 pulsars (NGC 6440A-H), this cluster also hosts seven probable quiescent low-mass X-ray binaries \citep{Heinke+2003} and it was the first cluster in which two luminous transient LMXBs were found \citep[both showing millisecond pulsations;][]{Heinke+2010}. A third transient X-ray binary was recently found in Terzan 5 \citep{2014ApJ_Bahramian+}. 
NGC 6440 has a relatively high interaction rate per binary, $\gamma \propto \rho_{\rm c} / \nu $, $\rho_{\rm c}$ is the density of the cluster, and $\nu$ is the velocity dispersion. Terzan 5 and NGC 6440 are among the most massive and densest clusters of the non-core collapsed GCs (see Table 1 from \citealt{Verbunt+Freire2014}). These clusters with `intermediate' $\gamma$ are expected to host a mix of isolated and binary pulsars, like in the case of NGC 6440 with the exact same number of isolated and binary pulsars known to date. The peculiarity of NGC 6440 is its large production of binary systems with low mass companions. This can be understood primarily as a consequence of the high density of the cluster and the large $\gamma$ value for a non-core-collapsed cluster. \citet{Krolik+1984} suggest that in clusters with high central density isolated NSs or NSs with a very low mass companions are produced by direct collision between main-sequence stars and NSs. Also, \citet{Ivanova+2008} find that the formation of low-mass X-ray binaries has a 50\% contribution by binary exchange, physical collisions with giants and tidal capture in high density clusters. Additionally, in these clusters with intermediate or high encounter rates per binary, the formation of a new LMXB system could be possible via ``secondary'' exchange encounters as a result of an already recycled pulsar acquiring a main sequence star companion \citep{Verbunt+Freire2014}. 

NGC 6440 has the characteristics one would expect for a GC with an intermediate to high $\gamma$: there is an apparently young pulsar, NGC 6440A, with a characteristic age $\sim 10^3$ times smaller than the age of the cluster, there is one binary system with a high eccentricity (similar to what is seen in clusters like Terzan 5 and M28), it also hosts two pulsars that lie more than 4 core radius from the centre, NGC 6440C and NGC 6440D. But no pulsars have been found further than the half mass radius, like in the case of other core-collapsed clusters (e.g. NGC 6752; \citealt{D'Amico+2002, Corongiu+2006}). 
For all these reasons it is not surprising that the pulsar with the lightest companion known so far in a GC is found in this cluster.
It is important to note that although Terzan 5 and NGC 6440 have approximately the same encounter rate and the same encounter per binary rate, they have different pulsar populations and the particular internal distribution of spin periods is different. The past dynamical history of the cluster must have had some influence on these.



\section{Conclusions}
\label{s:Conclusions}

Two new pulsars have been discovered in NGC 6440 using MeerKAT: one of them is isolated, NGC 6440G, and the other one, NGC 6440H, is a Black widow which does not display eclipses in the 962-1605\,MHz band.The companion to NGC 6440H is also the lightest mass pulsar companion so far known in a GC, with a minimum mass of $0.006$\,\msun. 
The two new additions maintain the previously observed equal ratio of isolated and binary pulsars in NGC6440 (now 4:4), and contributes to the pulsar discoveries in GCs using MeerKAT. 

The high-cadence of the complementary MeerTime and TRAPUM observations, as well as the large bandwidth of MeerKAT also enabled us to study in detail the known pulsars NGC 6440C and NGC 6440D. On the basis of the new data, we could reject the existence of planets in simple configurations around the isolated pulsar NGC 6440C. Further studies of the profile variability are warranted to gather more information about the systematics found in the timing residuals of this pulsar. Our analysis of the eclipses of the Redback pulsar NGC 6440D at two different frequency bands revealed a clear frequency dependence, with longer and asymmetric eclipses occurring at the lower frequencies (962-1283\,MHz).

We note that the precise timing parameters obtained for both NGC 6440G and NGC 6440H would not have been possible without the availability of the archival GBT data that enabled us to derive a $\sim 15$\,years long timing solution. Moreover, the tiling capabilities of the TRAPUM back-end have helped us to obtain a better localisation for the very faint and relatively steep spectrum ($\alpha \approx -1.76$) pulsar NGC 6440G and thus improve the timing solution.

Additional observations of NGC 6440 and of a large set of other GCs are being collected as part of the TRAPUM GC pulsar survey, which exploits the full MeerKAT array.

\section*{Acknowledgements}
The MeerKAT telescope is operated by the South African Radio Astronomy Observatory, which is a facility of the National Research Foundation, an agency of the Department of Science and Innovation. SARAO acknowledges the ongoing advice and calibration of GPS systems by the National Metrology Institute of South Africa (NMISA) and the time space reference systems department of the Paris Observatory. MeerTime data is housed on the OzSTAR supercomputer at Swinburne University of Technology. The OzSTAR program receives funding in part from the Astronomy National Collaborative Research Infrastructure Strategy (NCRIS) allocation provided by the Australian Government. 
The Green Bank Observatory is a facility of the National Science Foundation operated under cooperative agreement by Associated Universities, Inc.
The National Radio Astronomy Observatory (NRAO) is a facility of the National Science Foundation operated under cooperative agreement by Associated Universities, Inc. 
The authors also acknowledge MPIfR funding to contribute to MeerTime infrastructure. TRAPUM observations used the FBFUSE and APSUSE computing clusters for data acquisition, storage and analysis. These clusters were funded and installed by the Max-Planck-Institut für Radioastronomie and the Max-Planck-Gesellschaft. 
LV acknowledges financial support from the Dean’s Doctoral Scholar Award from the University of Manchester. 

BWS and MCB acknowledge funding from the European Research Council (ERC) under the European Union’s Horizon 2020 research and innovation programme (grant agreement No. 694745)

EDB, MK, VVK, AR, PCCF, DJC, WC, and APa, acknowledge continuing valuable support from the Max-Planck Society. 

SMR is a CIFAR Fellow and is supported by the NSF Physics Frontiers Center awards 1430284 and 2020265.

AR, APo and MBu gratefully acknowledge financial support by the research grant ``iPeska'' (P.I. Andrea Possenti) funded under the INAF national call Prin-SKA/CTA approved with the Presidential Decree 70/2016.
AR, APo and MBu also acknowledge the support from the Ministero degli Affari Esteri e della Cooperazione Internazionale - Direzione Generale per la Promozione del Sistema Paese - Progetto di Grande Rilevanza ZA18GR02.
ICN acknowledges support from the STFC project grant ST/T506291/1.

MED acknowledges support from the National Science Foundation (NSF) Physics Frontier Center award 1430284, and from the Naval Research Laboratory by NASA under contract S-15633Y.

Pulsar research at UBC is supported by an NSERC Discovery Grant and by the Canadian Institute for Advanced Research.

J.W.T.H. acknowledges funding from an NWO Vici grant ("AstroFlash"; VI.C.192.045).
We thank the anonymous referee for the helpful comments, and constructive remarks on this manuscript.

\section*{Data Availability}
The data that support the findings of this article will be shared on reasonable request to the MeerTime and TRAPUM collaborations.



\bibliographystyle{mnras}
\bibliography{DT_NGC6440} 

\bsp	
\label{lastpage}
\end{document}